\documentclass[conference]{IEEEtran}
\IEEEoverridecommandlockouts
\usepackage{cite}
\usepackage{amsmath,amssymb,amsfonts}
\usepackage{algorithmic}
\usepackage{graphicx}
\usepackage{textcomp}
\usepackage{subfig}
\usepackage{url}

\usepackage{xcolor}
\def\BibTeX{{\rm B\kern-.05em{\sc i\kern-.025em b}\kern-.08em
    T\kern-.1667em\lower.7ex\hbox{E}\kern-.125emX}}
\begin{document}

\title{How Much Does It Hurt: A Deep Learning Framework for Chronic Pain Score Assessment\\
}

\author{\IEEEauthorblockN{Yun Zhao\IEEEauthorrefmark{1},
Franklin Ly\IEEEauthorrefmark{3}, Qinghang Hong\IEEEauthorrefmark{1}, Zhuowei Cheng\IEEEauthorrefmark{1}, Tyler Santander\IEEEauthorrefmark{4}, \\Henry T. Yang\IEEEauthorrefmark{3}, Paul K. Hansma\IEEEauthorrefmark{2}, Linda Petzold\IEEEauthorrefmark{1}}
\IEEEauthorblockA{\IEEEauthorrefmark{1}Department of Computer Science, University of California, Santa Barbara, CA, USA\\
\IEEEauthorrefmark{2}Department of Physics, University of California, Santa Barbara, CA, USA\\
\IEEEauthorrefmark{3}Department of Mechanical Engineering, University of California, Santa Barbara, CA, USA\\
\IEEEauthorrefmark{4}Department of Psychological \& Brain Sciences, University of California, Santa Barbara, CA, USA\\
yunzhao@cs.ucsb.edu}}

\maketitle

\begin{abstract}
Chronic pain is defined as pain that lasts or recurs for more than 3 to 6 months, often long after the injury or illness that initially caused the pain has healed. The “gold standard” for chronic pain assessment remains self report and clinical assessment via a biopsychosocial interview, since there has been no device that can measure it. A device to measure pain would be useful not only for clinical assessment, but potentially also as a biofeedback device leading to pain reduction. In this paper we propose an end-to-end deep learning framework for chronic pain score assessment. Our deep learning framework splits the long time-course data samples into shorter sequences, and uses Consensus Prediction to classify the results. We evaluate the performance of our framework on two chronic pain score datasets collected from two iterations of prototype Pain Meters that we have developed to help chronic pain subjects better understand their health condition.

\end{abstract}


\section{Introduction}
Deep learning models have achieved remarkable success in computer vision~\cite{Imagenet}, natural language processing~\cite{NLP}, speech recognition~\cite{Speech} and the game of Go~\cite{Alphago}. Recently there has been increasing interest in applying deep learning for end-to-end health data analysis~\cite{avati2018improving}. However, e-health data analysis is even more challenging since well-being data can be affected by many factors, for example individual differences, measurement errors in data collection and missing data.

Chronic pain is described as persistent or recurrent pain that lasts for at least 3 to 6 months~\cite{dahlhamer2018prevalence}. According to the 2016 National Health Interview Survey (NHIS), roughly 20.4\% (50.0 million) of U.S. adults suffer from chronic pain. Chronic pain affects individuals, their families, and society, and results in complications harming both physical and mental health. The economic costs of chronic pain and pain-related disability in the United States cannot be overstated. One influential report~\cite{gaskin2012economic} conservatively estimated an annual toll of \$560-\$650 billion dollars—far exceeding the costs of cardiovascular disease, cancer, and diabetes. Therefore, identifying the score of chronic pain is of significant value to reduce further complications. 

Neurophysiological signals have been used to quantify pain~\cite{van2019neuroimaging, reddan2019brain, reddan2018modeling}. In the last decade, there has been some progress towards discovering the neurobiological substrates of pain. However, none of those methods is low-cost or easy-to-use. Clearly, there is an unmet need for a low-cost, easy-to-use Pain Meter. According to recent research results~\cite{cowen2015assessing,yang2018postoperative,prichep2018exploration}, classical physiological measurements can effectively quantify pain. Inexpensive technology is now available to measure relevant physiological features. Taken together, it is clear that objectively quantifying pain is possible. We thus investigated a number of inexpensive commercial sensors capable of measuring physiological variables for pain score assessment. We have thus far built prototype Pain Meters that offer the immense potential to revolutionize pain treatment and the development of therapeutics.

To this end, we collected two new chronic pain datasets, using prototype Pain Meter 1 and 2, respectively. For Dataset 1, our subject has been suffering chronic pain for more than 10 years. Neck and shoulder pain were causing her difficulty to perform daily activities.  For Dataset 2, we recruited chronic pain subjects from our local community. All subjects signed an informed consent form according to a protocol approved by a Human Subjects Committee.
Our chronic pain datasets are characterized by the following unique properties:
First, we use Photoplethysmography (PPG)~\cite{liu2012review}, which is low cost and comfortable for patients, to collect pulse signals. Secondly, we also include several temperature signals and Galvanic Skin Response(GSR) signals to detect the chronic pain symptoms like nervousness. For Dataset 2, we also use accelerometers and gyros to detect movements.

\begin{figure*}[ht]
  \includegraphics[width=\linewidth]{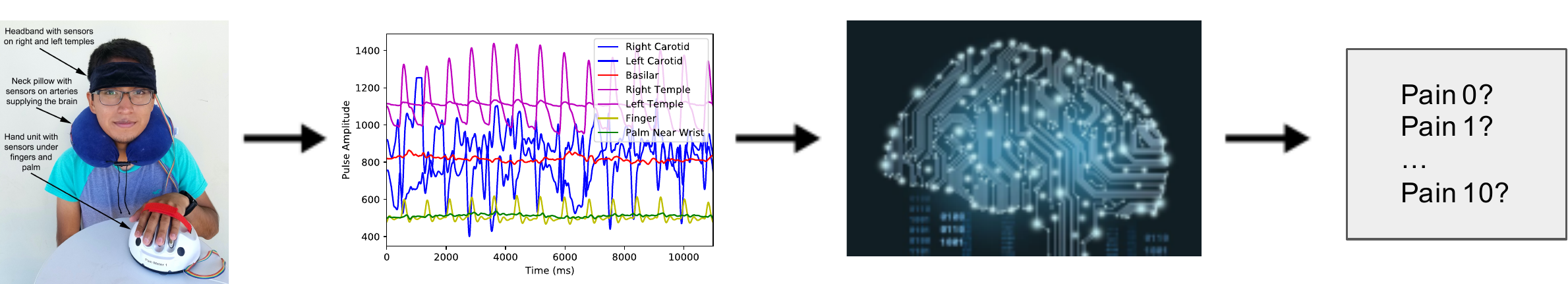}
  \caption{The workflow for chronic pain score assessment. The brain image is from Google.}
  \label{framework}
\end{figure*}

Given the pain score datasets, we introduce the task of chronic pain score prediction, which is to train an end-to-end oridinal classifier to accurately predict pain score. The illustration of our workflow is shown in Fig.~\ref{framework}. An accurate chronic pain score assessment will
\begin{enumerate}

\item	Facilitate the development of new therapies both in the laboratory and in Phase II and Phase III clinical trials.
\item	Make it possible for physicians to quantify the effects of existing therapies on individual patients.
\item	Minimize the harm caused by diagnostic delays and the under/overtreatment of pain due to the influence of gender, race, or age.

\item	Allow a patient to decide, with objective personal data, whether current treatments are effective in his/her quest to reduce chronic pain.

\item  Serve as a biofeedback device for chronic pain subjects. The unconscious mind can learn to control things it can monitor~\cite{bargh2008unconscious}. If people are enabled to accurately monitor their chronic pain score with a Pain Meter, their unconscious mind can figure out how to decrease the chronic pain score. It is trained and rewarded by the tiny decreases in pain score that are accurately and continuously monitored. 
\end{enumerate}

Our main contributions are threefold:

\begin{enumerate}
\item We propose a deep learning ordinal regression framework for chronic pain score assessment. To the best of our knowledge, this is the first paper to use deep learning for chronic pain score assessment.
\item We collect two new chronic pain datasets using our prototype Pain Meters to predict the score of chronic pain.
\item We split the long recordings into smaller slices for training, which not only eases the burden on GPU memory but also provides many training samples for deep learning models. We define Consensus Prediction as the majority voting result of the sampled short slices for testing, since not all of the short slices can be expected to contain enough useful information. 
\end{enumerate}

The remainder of this paper is organized as follows. Section~\ref{Related_work} discusses related work. Section~\ref{Data} describes the Pain Meters we used for data collection and introduces the classification problem. Section~\ref{Model} delineates the methodology. Section~\ref{Experiment} introduces the experimental setup for comparison of our results to those obtained via multilayer perceptron and logistic regression. Results and case studies are described in Section~\ref{Results}. Section~\ref{optimize} shows that our deep learning framework can also be used to provide feedback to improve the design of the Pain Meter. Section~\ref{Discuss} is the Discussion.

\section{Related Work}\label{Related_work}
\subsection{Deep Learning for E-Heath}
Since the emergence of deep learning in e-health, more and more researchers have been implementing deep learning models for medicine, aiming to improve health care~\cite{avati2018improving}, classify diseases~\cite{kanawade2019photoplethysmography}, and prevent misdiagnosis~\cite{poplin2018prediction}. More specifically, the prediction of medical events has been popular, including the prediction of death rates~\cite{darabi2018forecasting}, prescriptions~\cite{jin2018treatment}, and successful extubation~\cite{hsieh2018artificial}. Although many researchers have applied deep learning to e-health, to the best of our knowledge no researcher has applied deep learning from pulse signals to the prediction of chronic pain scores~\cite{miotto2017deep}. 
There are image based models for automatic estimation of pain~\cite{rodriguez2017deep,liu2017deepfacelift}. Deep Pain~\cite{rodriguez2017deep} uses long short term-memory (LSTM) and analyzes images of  facial expression. However, images alone are not reliable since images alone may not contain enough relevant information for pain score prediction. Many factors affect pain score, including stress and mood, which might not be fully captured by image based methods. Physio-based models have been developed for assessments of physiological pain~\cite{chu2017physiological} but these have not been applied to chronic pain. Motivated by this need, we propose a Convolutional Neural Network (CNN)-based framework that uses physiological signals to assess chronic pain scores.

\subsection{Chronic Pain and Traditional Machine Learning}

In the context of pain assessment research, physiologically-based pain has been the main focus for many pain researchers ~\cite{lotsch2018machine}. Chronic pain, on the other hand, is prolonged, lasting anywhere from months to years~\cite{pitcher2019prevalence}. In the past, traditional machine learning methods have been applied to e-health, but this has required extensive feature extraction~\cite{johnson2016machine}. Human feature extraction has many disadvantages. For example,  it is costly and can result in the “loss of data interpretability”~\cite{hira2015review}. Random Forest has been used to monitor the nociception (perception of pain) level, which requires feature extraction~\cite{ben2013monitoring}. Physiological parameters like heart rate, heart rate variability, plethysmograph wave amplitude, skin conductance level, number of skin conductance fluctuations, and their time derivatives are extracted for prediction. In contrast, our end-to-end deep learning framework requires no feature extraction and little data preprocessing.

\subsection{Physiological Sensors}
Photoplethysmography (PPG) is a widely used non-invasive method for measuring pain intensity, in which changes in blood volume and light absorption are detected~\cite{liu2012review}. This type of technology is commonly used because it is simple and can easily collect data, while being cost-efficient and accessible~\cite{rubins2019imaging}. A PPG device has also been developed to monitor respiratory and heart rates of infants, and this has been proven to perform well~\cite{johansson1999monitoring}. PPG is favored not only for clinical use, but also for home use as a biofeedback device. In addition, it is more comfortable for patients because it does not involve gel and electrodes contacting their skin~\cite{bonissi2013preliminary}. Thus, in our prototype Pain Meter, we use PPG~\cite{PPG_link} to collect pulse recordings. We also included axis accelerometer gyroscope modules (MPU-6050), force sensitive resistors to measure the forces that cause low frequency motion (DF9-40), and a GSR sensor (ZIYUN Grove GSR sensor).

\section{Data Collection and Classification}\label{Data}
\subsection{Prototype Pain Meter}
\begin{figure}[ht]
  \includegraphics[width=\linewidth]{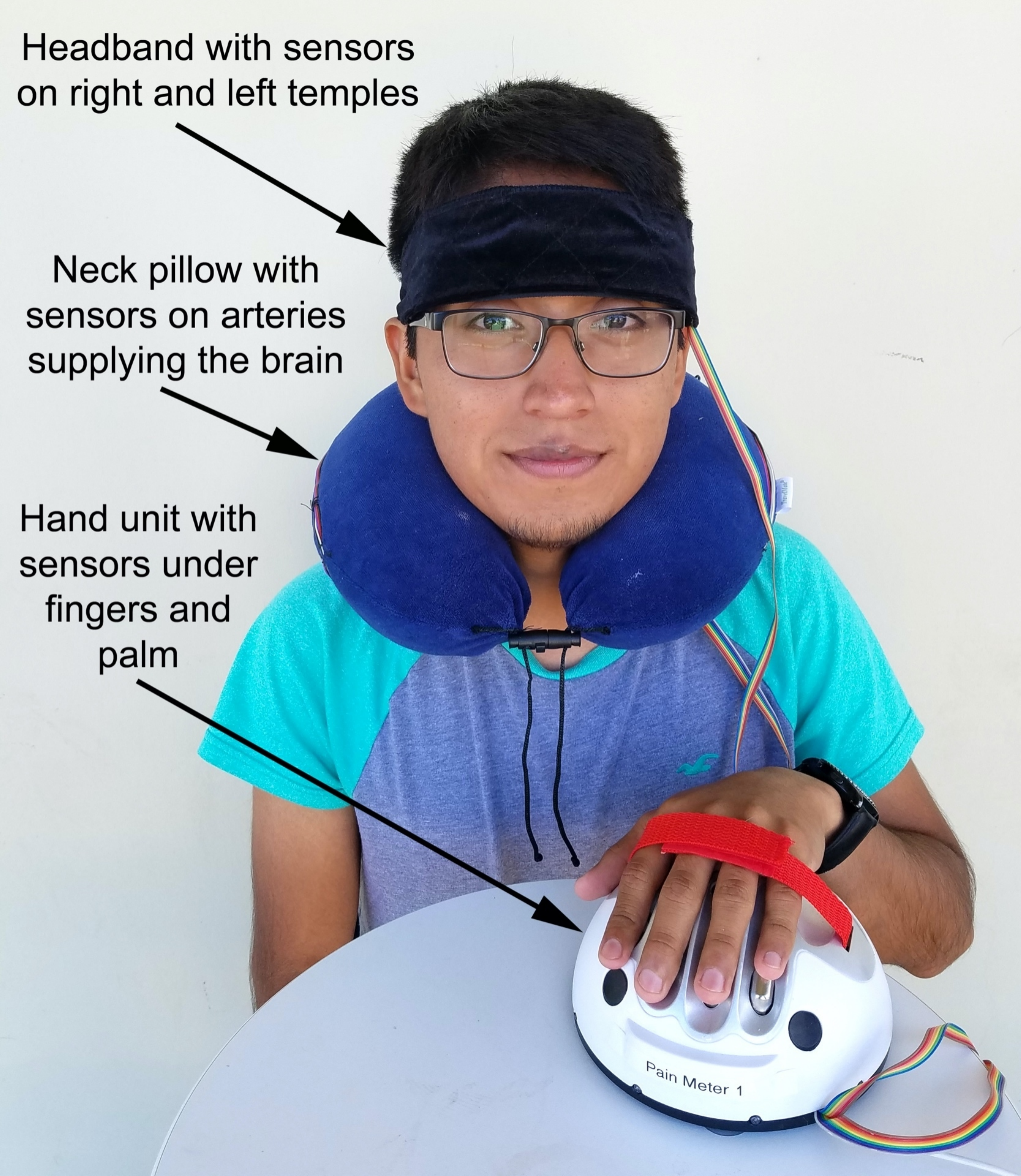}
  \caption{Pain Meter 1 sensed temperature, pulse and GSR, but it did not directly sense motion.}
  \label{meter1}
\end{figure}

Fig.~\ref{meter1} shows the device we used for collecting Dataset 1. Pain Meter 1 contains: 1) two PPG pulse sensors held to the temples via a headband, three at the three arteries supplying blood to the brain mounted in a neck pillow, and two at the fingertip and palm, 2) temperature sensors at each location of the PPG pulse sensors, and 3) GSR sensors embedded in the block on which the hand rests. These provide a total of 15 signals, recorded in Dataset 1.  

\begin{figure}[ht]
  \includegraphics[width=\linewidth]{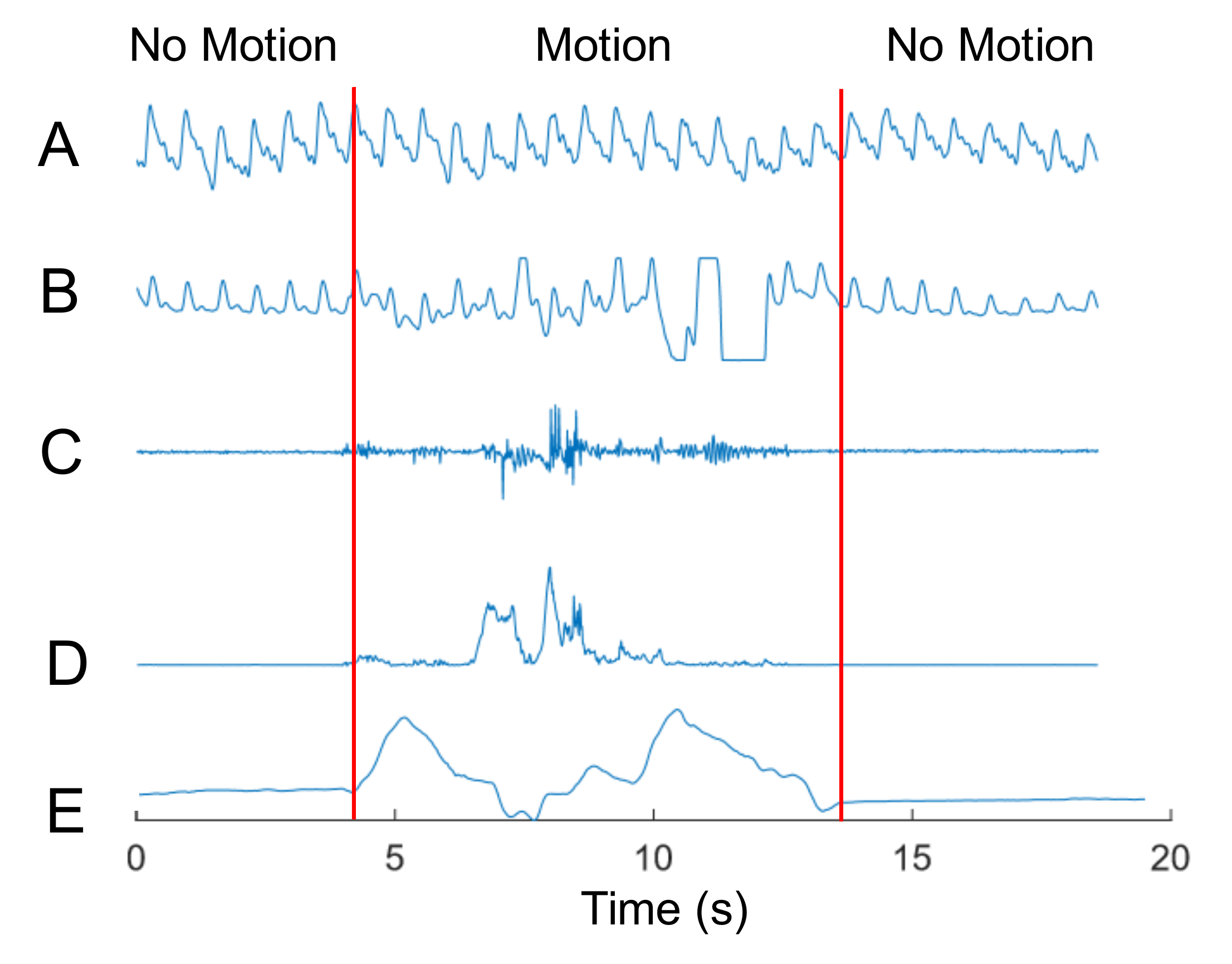}
  \caption{Motion affects the pulse sensor signals, which are based on measuring the intensity of light reflected from the skin. Part A shows a pulse sensor at the carotid artery, far from motion. Part B shows a pulse sensor at the finger. Motion is generated by flexing the wrist and is measured in part C by a wrist acceleration sensor. Part D shows data from wrist gyro sensor and part E shows data from a wrist force sensor. Note that the pulse sensor on the finger (sensor B), which is closer to the motion, is strongly affected by the motion, while the pulse sensor on the carotid artery (sensor A) is not.}
  \label{why}
\end{figure}

\begin{figure}[ht]
  \includegraphics[width=\linewidth]{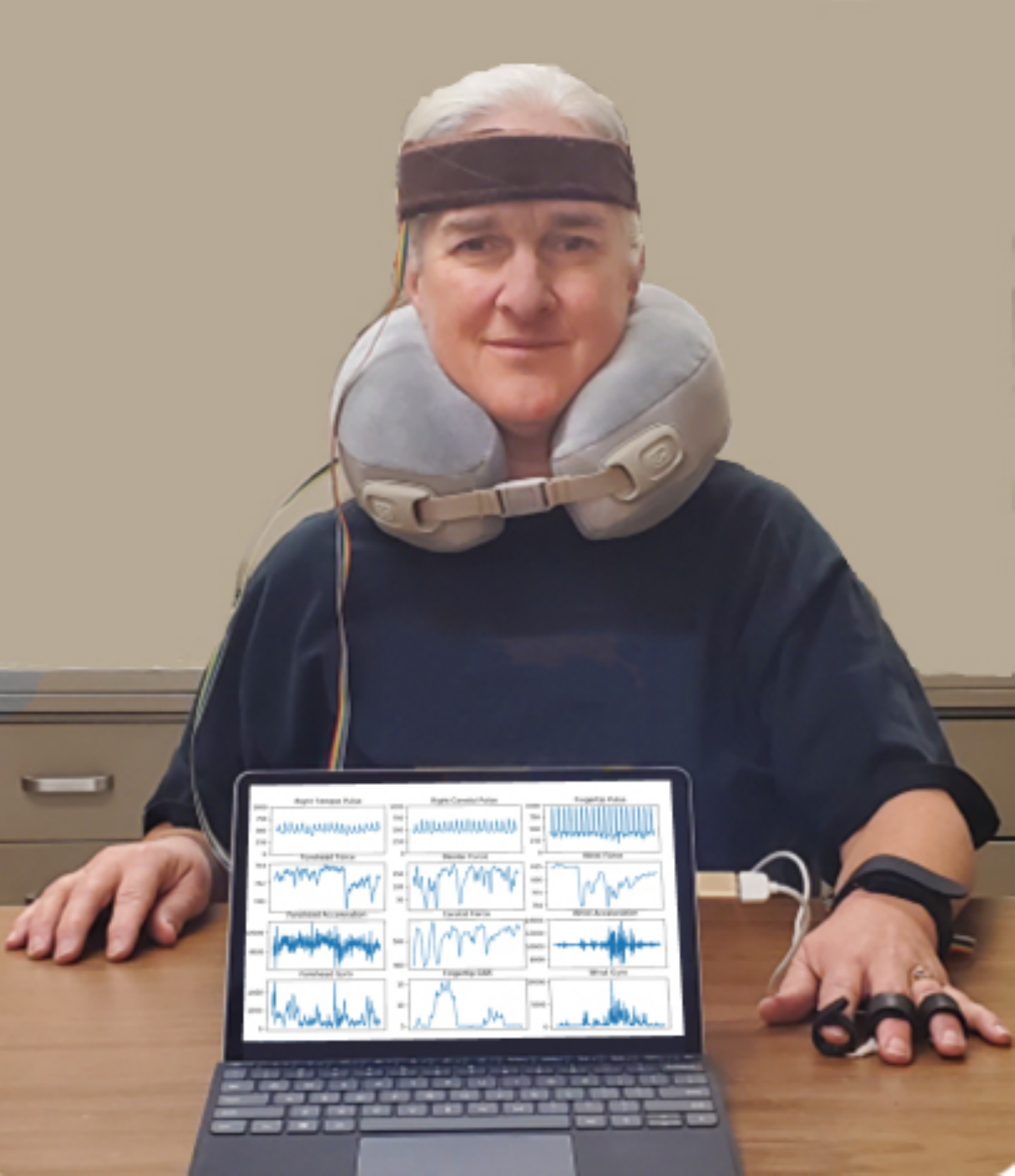}
  \caption{Pain Meter 2 consists of: 1) PPG pulse sensors in a headband, in a neck band, and on the fingertip, 2) temperature sensors on the neck and fingertip, 3) 3-axis accelerometers and 3-axis gyros in the head band and wrist band, 4) force sensors on the forehead, back of neck, side of neck, and wrist band, and 5) GSR sensors between the middle and ring fingers.}
  \label{meter2}
\end{figure}

As is shown in Fig.~\ref{why}, it turned out that the PPG sensors were sensing more than pulse. They were also sensing subtle motion. Motivated by these phenomena, we added actual motion sensors in our Pain Meter 2. As is shown in Fig.~\ref{meter2}, Pain Meter 2 contains: 1) PPG pulse sensors in a headband, in a neck band, and on the fingertip, 2) temperature sensors on the neck and fingertip, 3) 3-axis accelerometers and 3-axis gyros in the head band and wrist band, 4) force sensors on the forehead, back of neck, side of neck, and wrist band, and 5) GSR sensors between the middle and ring fingers. \textcolor{black}{These provide a total of 25 signals, recorded in Dataset 2.}

For both Pain Meters, a Teensy 3.6 microcontroller with a 32-bit 180 MHz ARM Cortex-M4 processor is used to sample all signals every 15ms. A data acquisition program was designed in MegunoLink Pro for Pain Meter 1 data and a customized data acquisition program was written in Python for Pain Meter 2 data. 

\subsection{Data Collection}

Using Pain Meter 1 from Fig.~\ref{meter1}, our chronic pain Dataset 1 was collected from one subject who self-reported the pain score. The subject was asked to fix the headband and neck pillow into a comfortable position. Once secured, real-time plots of the pulse and temperature data were viewed to verify that the head and neck sensors were collecting accurate and reliable signals for the subject. Minor adjustments to the sensor positions can be made if needed. After adjustments were made, the subject placed the left hand on the module to read the hand signals. A final verification step viewing all plots of data was performed. Then the subject was asked to close their eyes and relax before the recording begins. After 10 minutes, the recording was ended. Each recording was taken on a different day at the same time in the afternoon, the same temperature and the same environment brightness. During recordings with Pain Meter 1 we noticed that some of the pulse signals became erratic compared to other pulse signals and that this erratic behavior seemed to correlate with pain. We hypothesized that this was due to subtle movement~\cite{zuzarte2019quantifying} and constructed Pain Meter 2 (Fig.~\ref{meter2}) to have motion sensors. With the addition of motion sensors we could confirm this hypothesis (Fig.~\ref{why}) and measure the motion directly.

\begin{figure}[ht]
  \includegraphics[width=\linewidth]{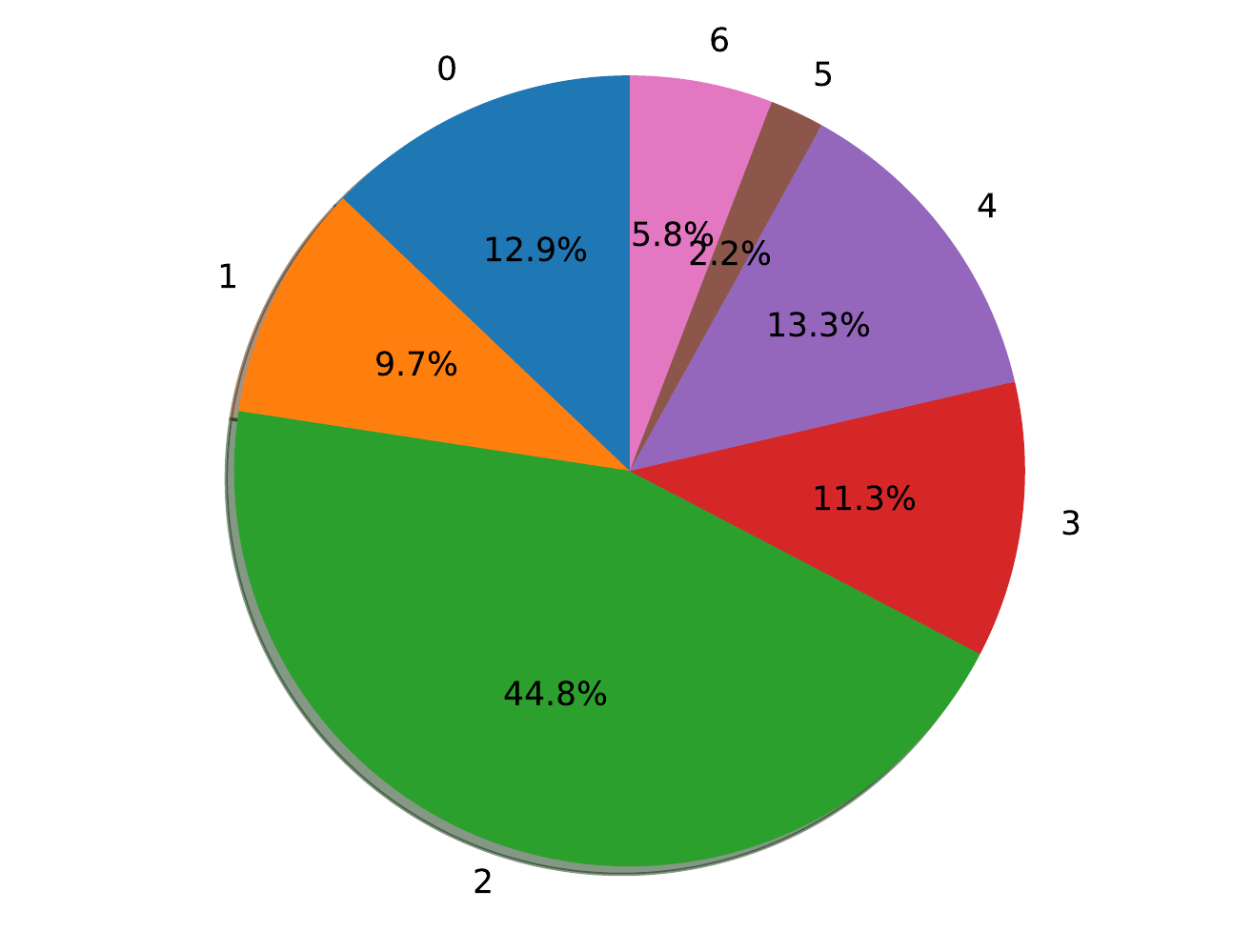}
  \caption{Distribution of pain scores in Dataset 2. The numbers outside the pie chart are the corresponding pain scores and the percentages of time for which each pain score is reported are shown in the pie chart.}
  \label{fig:statistics}
\end{figure}

Pain score distributions for the 2-class Dataset 1 and 7-class Dataset 2 are shown in Table~\ref{data_summaries} and Fig~\ref{fig:statistics}, respectively. Each recording has a length of 10 minutes, with signals sampled every 15 milliseconds. We have 4 recordings from one subject in Dataset 1 and 62 recordings from 20 subjects in Dataset 2. We divide each 10-minute recording into ten mutually exclusive 1-minute samples.

\newcommand{\tabincell}[2]{\begin{tabular}{@{}#1@{}}#2\end{tabular}}  

\begin{table}
\caption{Data statistics for Dataset 1}\label{data_summaries}
\begin{center}
\begin{tabular}{|l|l|l|}
\hline
Data  & Classes &  \tabincell{c}{Pain score\\ distribution }\\
\hline
Chronic Pain  & 2 & 1 : 1\\
\hline
\end{tabular}
\end{center}
\end{table}

\subsection{Classification Problem}
The classification problem for chronic pain assessment is to classify a Pain Meter dataset of time sequences into pain scores on a scale of 0 for no pain to 10 for the worst pain possible. The model is first trained on Dataset 1 from a chronic pain subject who self reports her pain score for each of the datasets on a 0 to 10 scale. Note that seldom has our subject reported a pain score higher than 2 in Dataset 1. Thus in Dataset 1, we use pain scores 1 and 2. In Dataset 2, we have pain scores from 0 to 6, which formulates to be a 7 class classification problem. Our goal is to use the trained deep learning model to accurately classify chronic pain datasets into the self reported pain scores. 

\section{Methodology}\label{Model}

\begin{figure*}[ht]
  \includegraphics[width=\linewidth]{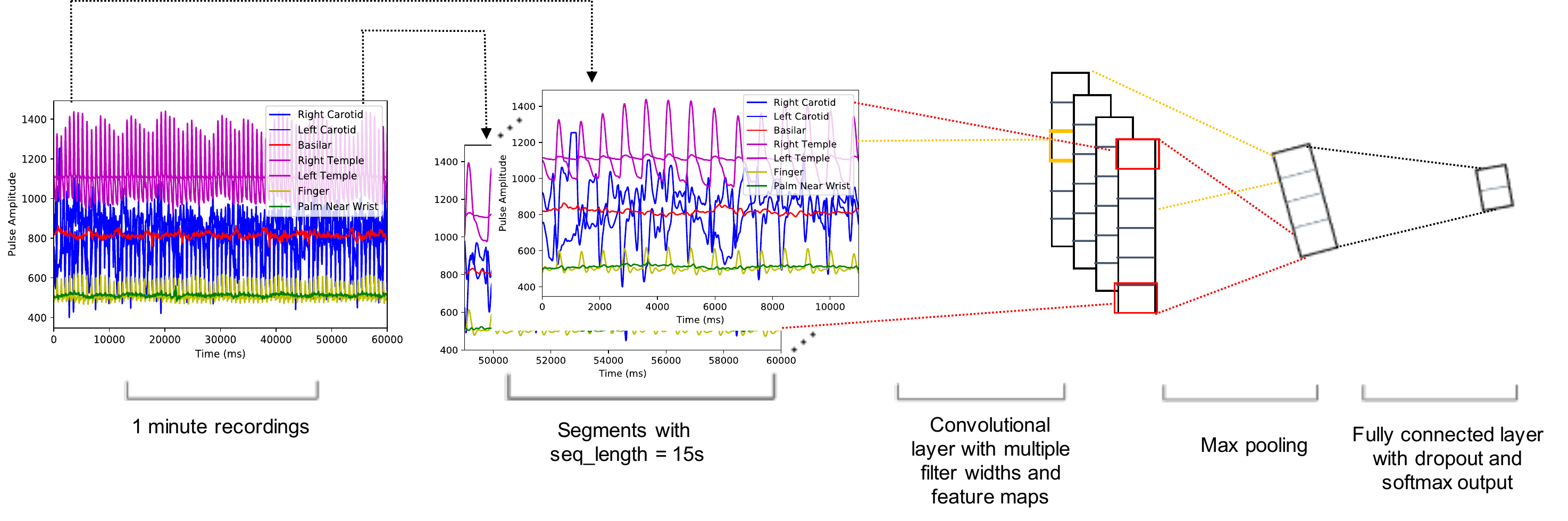}
  \caption{Model architecture: 1-minute samples of the voltages measured on the physiological sensors are collected from our prototype Pain Meter. Segments with seq\_length = 15 s of these samples are individually classified. These individual classifications are conducted for Consensus Prediction in Pain score recordings.}
  \label{model}
\end{figure*}

The model architecture, shown in Fig.~\ref{model}, consists of convolution-pooling layers followed by fully connected layers.
To learn temporal and correlation features, the convolution is performed on both the time and sensor dimensions. We split the 1-minute samples into smaller slices with length of seq\_length. Sensor recording signals with a dimension of (number of sensors (N) $\times$ seq\_length) serve as input $x$ for the neural network. A convolution operation involves a filter $w \in \mathbb{R}^{st}$, which is applied to a window of s sensors and t samples to produce a new feature. For example, a feature $f_{i,j},(0 \leq i \leq N - s + 1, 0 \leq j \leq seq\_length - t + 1)$ is generated from a window size $(s,t)$ of the sensor signals:
\begin{equation}
f_{i,j} = ReLU (w · x_{i:i+s-1,j:j+t-1} + b),
\end{equation}
where $b \in \mathbb{R}$ is a bias term. This filter is applied to each possible window of the voltage signals to produce a feature map:
\begin{equation}
\resizebox{.9\columnwidth}!{
$f=
  \left[ {\begin{array}{cccc}
   f_{1,1} & f_{1,2} & . . . & f_{1,seq\_length-t+1} \\
   f_{2,1} & f_{2,2} & . . . & f_{2,seq\_length-t+1}\\
   . . . & . . . & . . . & . . .\\
   f_{N-s+1,1} & f_{N-s+1,2} & . . . & f_{N-s+1,seq\_length-t+1}
  \end{array} } \right],$
  }
\end{equation}
with $f \in \mathbb{R}^{N-s+1,seq\_length-t+1}$. We then apply a max-pooling operation over the feature map and take the maximum value $m = \max{(f)}$ as the feature corresponding to this particular filter. The idea is to capture the most important feature, the one with the highest value, for each feature map. Our model uses multiple filters to obtain multiple features. These features form the penultimate layer and are passed to a fully connected softmax layer whose output is the probability distribution over different pain scores. We adjust the number of convolutional ReLU layers from 2 to 5, based on the choice of seq\_length. The prediction of our model, parameterized by $\boldsymbol{W}$,
is given by $\boldsymbol{p}(\hat{\boldsymbol{y}}|\boldsymbol{x},\boldsymbol{W})$.

Due  to  the  existence  of  the inherent ordering information in our chronic pain score, we apply ordinal  regression, a setting that bridges metric regression and classification, to predict chronic pain scores of ordinal scale. Compared to regular regression problems, these pain scores are discrete. These pain scores are also different from the labels of multiple classes in classification problems due to the existence of the ordering information. The cross entropy loss of our ordinal regression for input vector $\boldsymbol{x}_n$ is as follows,

\begin{equation}
\begin{aligned}
L_n(\boldsymbol{W}) = &(1 + |\frac{argmax(\boldsymbol{ p}(\boldsymbol{\hat{y}} |\boldsymbol{x}_n,\boldsymbol{W}) ) - \boldsymbol{y}}{C - 1}|)\\
&\times \frac{1}{C}(\sum_{i=1}^{C} - \boldsymbol{y}_i \cdot log(\boldsymbol{p}(\boldsymbol{\hat{y}} |\boldsymbol{x}_n,\boldsymbol{W}))),
\end{aligned}
\label{loss}
\end{equation}
where $\boldsymbol{y}$ represents the true pain scores, $\boldsymbol{p(\hat{y})}$ denotes the predicted probability vector with one value for each possible pain score, and $C$ is number of pain scores. Divided by ${C- 1}$, the absolute error $|argmax(\boldsymbol{ p}(\boldsymbol{\hat{y}} |\boldsymbol{x}_n,\boldsymbol{W}) ) - \boldsymbol{y}|$ is normalized between 0 and 1, with $C$ classes in total.  By multiplying $|\frac{argmax(\boldsymbol{ p}(\boldsymbol{\hat{y}} |\boldsymbol{x}_n,\boldsymbol{W}) ) - \boldsymbol{y}}{C - 1}|$, the normalized absolute error between prediction and ground truth, with cross entropy loss, we include the ordinal information in our loss function. We penalize more in our loss function if the absolute error between ground truth and prediction is larger. 

For testing, we define Consensus Prediction to measure the performance of predictions for the whole sensor signal sample. Consensus Prediction synthesizes results from multiple short slices by majority voting, which can significantly improve the prediction accuracy for a long sample. This is because not all of the short time slices can be expected to contain useful information for classification. 

We use Batch Normalization~\cite{BatchNorm} to accelerate training. For regularization, dropout~\cite{Dropout} and early stopping methods~\cite{Earlystop} are implemented to avoid overfitting. Dropout prevents co-adaptation of hidden units by randomly dropping out a proportion of the hidden units during backpropagation. Model training is ended when no improvement is seen during the last 100 validations. Softmax cross entropy loss is minimized with the Adam optimizer~\cite{Adam} for training.
Since both frequency information and correlation between sensors are captured through convolution filters, our CNN-based framework automatically deals with the features needed for classificaiton. We use grid search for hyperparameter tuning. The hyperparameters are described in Table~\ref{Hyp_tab}. 

\section{Experimental Setup}\label{Experiment}
We introduce two baselines: Multilayer Perceptron (MLP) and Logistic Regression, to compare with our proposed CNN framework for the two classification problems.
\subsection{CNN based Model and Consensus Prediction}
We implement a parallel processing framework that distributes the convolutional neural network into multiple ($N$) GPUs to ease the burden on GPU memory. Each GPU contains an entire copy of the deep learning model. We first split the training batch evenly into $N$ sub-batches. Each GPU processes only one of the sub-batches. Then we collect gradients from each replicate of the deep learning model, aggregate them together and update all the replicates. We train our CNN based framework with two NVIDIA GeForce GTX 1080s, each of which has a memory of 11178 MB.\\

\subsection{Multilayer Perceptron}

We use a four layer fully  connected network,  whose output is the probability distribution over two different classes, as a baseline.  We use a parallel processing framework similar to our CNN model implementation. We train our MLP on two NVIDIA FeForce GTX 1080s.

\subsection{Logistic Regression}

Since the signals are periodic, we extract individual voltage signal FFT ($x_1$) and pairwise sensor signal Pearson Correlation ($x_2$) as features ($X$) for a Logistic Regression classifier:

\begin{equation}
P(Y=1|X) =\frac{1}{1+e^{-(w_0+w_1x_1+w_2x_2)}},
\end{equation}
where $Y$ is the label for classification and $P$ is the probability of predicting $Y$ as label 1 (pain score 1). $w_0$, $w_1$ and $w_2$ are model parameters to be learned during training.

\subsection{Split Training and Testing}

Since each of our recordings is long enough to split into multiple informative samples and there exists different settings among different recordings, we implement two methods of splitting the training and testing data. We first divide each 10-minute recording into ten mutually exclusive 1-minute samples.\\

\begin{enumerate}

\item Considering  the  size  of  our  datasets in  prediction, we use 5-fold cross  validation.
For each of the 10-minute recordings, we use between $2 * (i - 1)$ and $2 * i$ minutes as testing, and the rest for training, for the $i$-th fold. This training/testing split
fits well with the scenario of practical use of pain score assessment. It is known that different subjects have different perceptions of pain. For real use, our pain meters need some self-calibration before getting accurate pain score readouts.\\

\item Leave-one-recording-out cross validation on all the recordings. Each recording is used once as a test set while the remaining recordings form the training set. We note that, due to differences in subjects' sensitivity to pain, different settings while recordings, and the size of our data, it is much more difficult to predict across subjects than within one recording. We report the result using this splitting method in Section~\ref{Discuss}.
\end{enumerate}

\begin{table}
\caption{Hyperparameters}\label{Hyp_tab}
\begin{center}
\begin{tabular}{|l|l|}
\hline
Hyperparameters  &   Value\\
\hline
Batch size  & 24 \\
Epoch   & 2000\\
Dropout rate   & 0.5\\
Seq\_length & 15 seconds\\
Learning rate & 0.5\\
\hline
\end{tabular}
\end{center}
\end{table}

\section{Results and Analysis}\label{Results}

Considering the size of our dataset in prediction, we use 5-fold cross validation and report the average results in this section. Note that Dataset 2 is unbalanced. We also report the confusion matrix for evaluation.

\begin{figure}[ht]
  \includegraphics[width=0.9\linewidth]{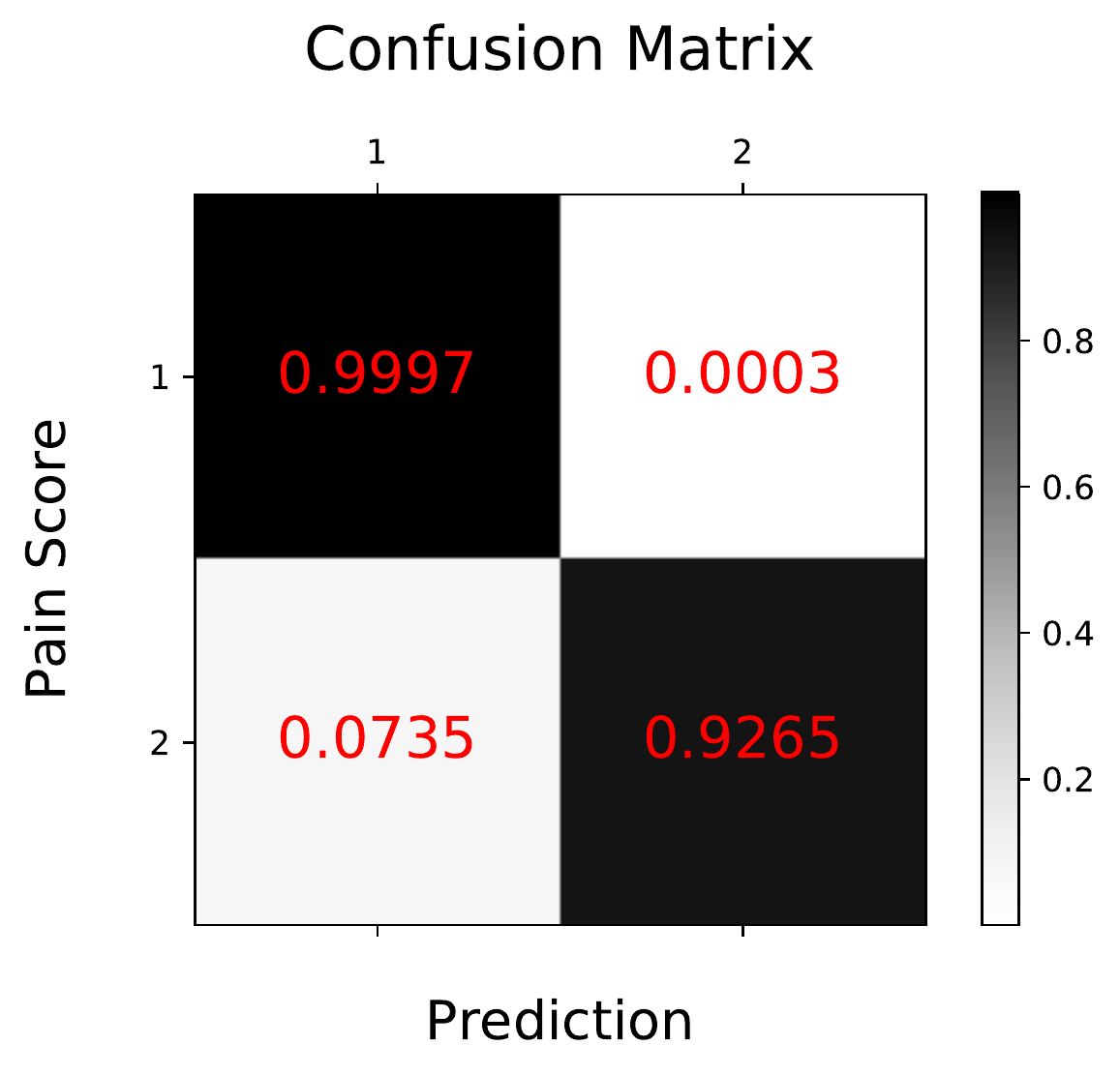}
  \caption{Confusion matrix for chronic pain score prediction in Dataset 1. }
  \label{Confusion_Matrix_1}
\end{figure}

\begin{figure}[ht]
  \includegraphics[width=\linewidth]{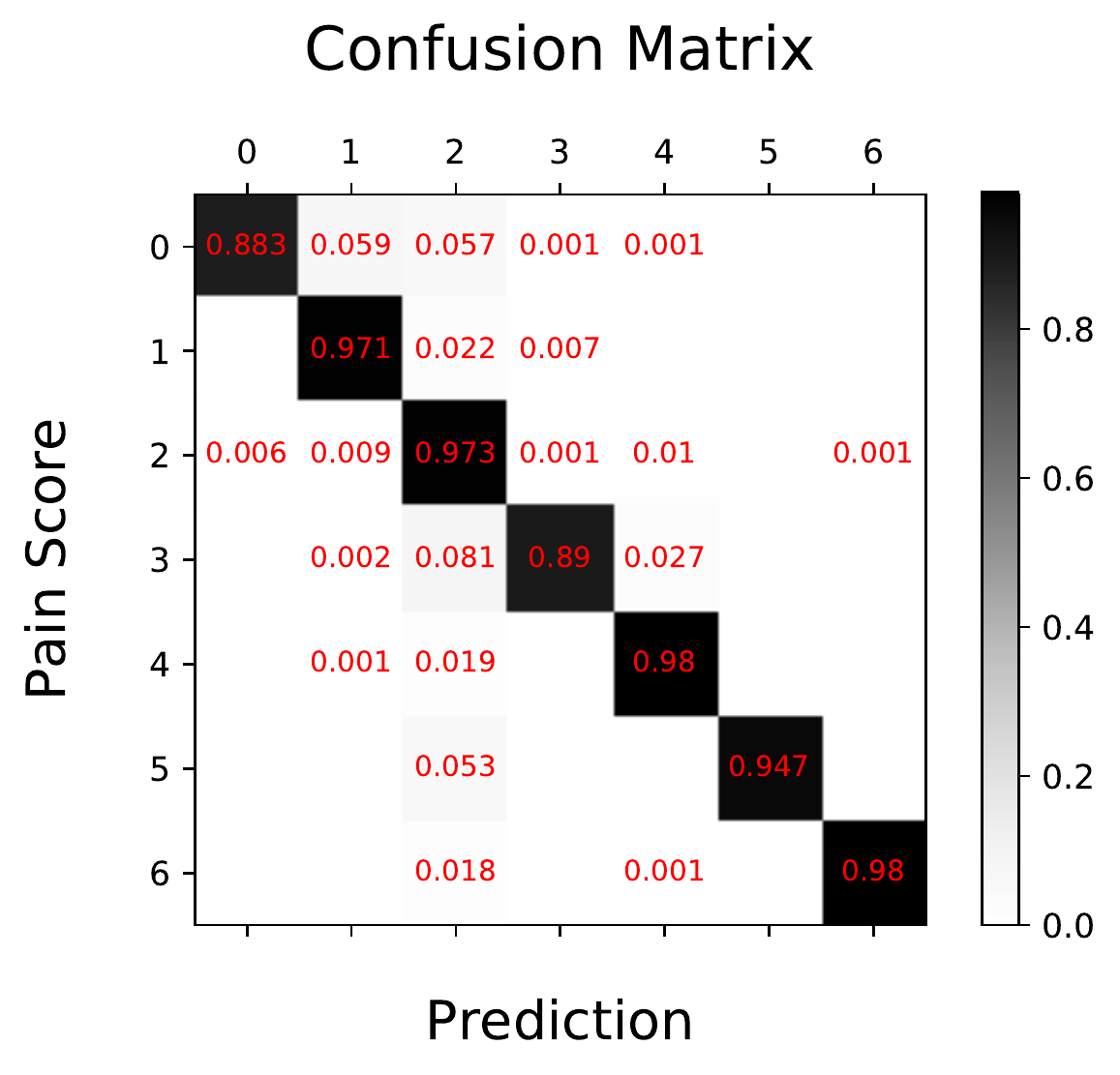}
  \caption{Confusion matrix for chronic pain score prediction in Dataset 2. Our model achieves high accuracy for each class.}
  \label{Confusion_Matrix_2}
\end{figure}

\begin{figure}[ht]
    \centering
    \subfloat{\includegraphics[width=0.23\textwidth]{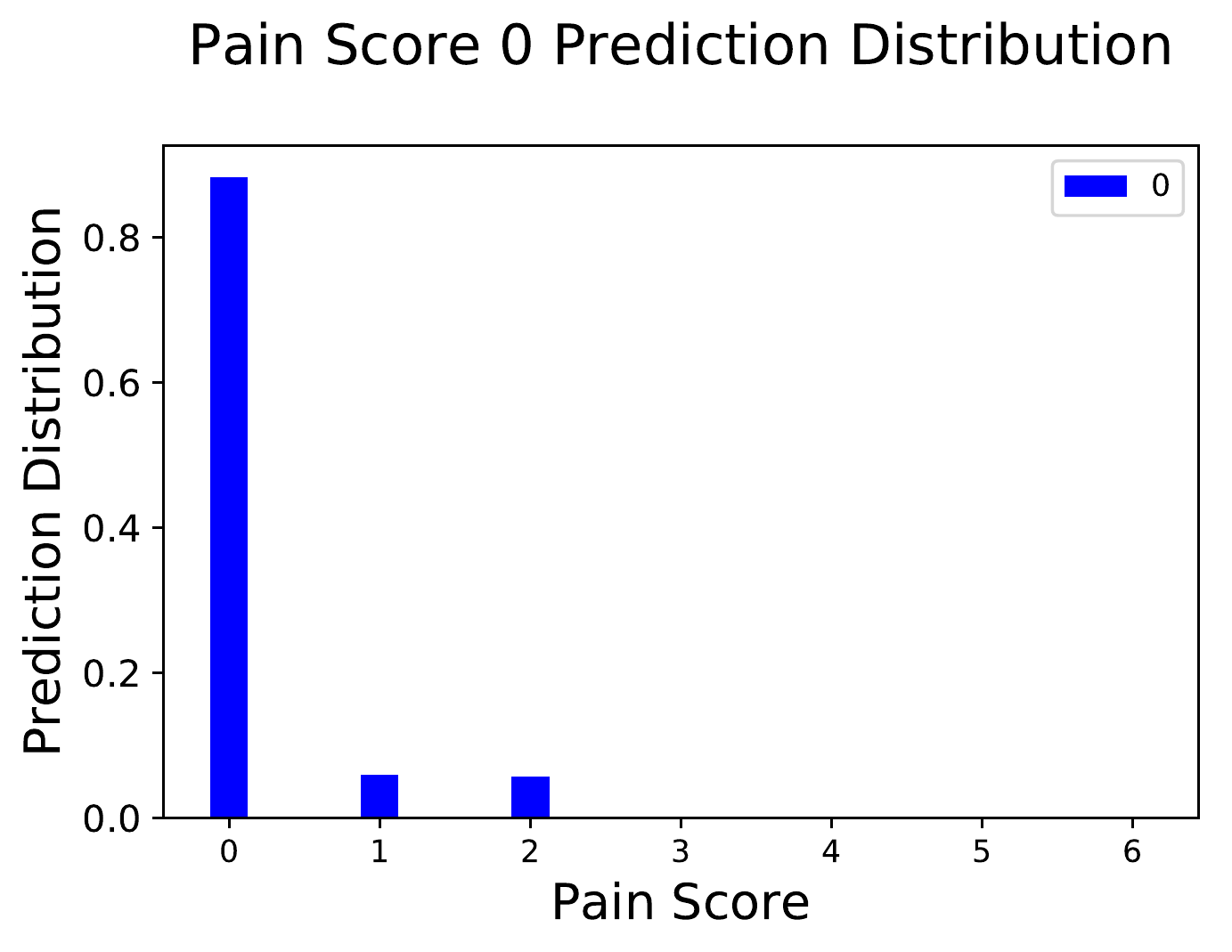}}
    \hfill
    \subfloat{\includegraphics[width=0.23\textwidth]{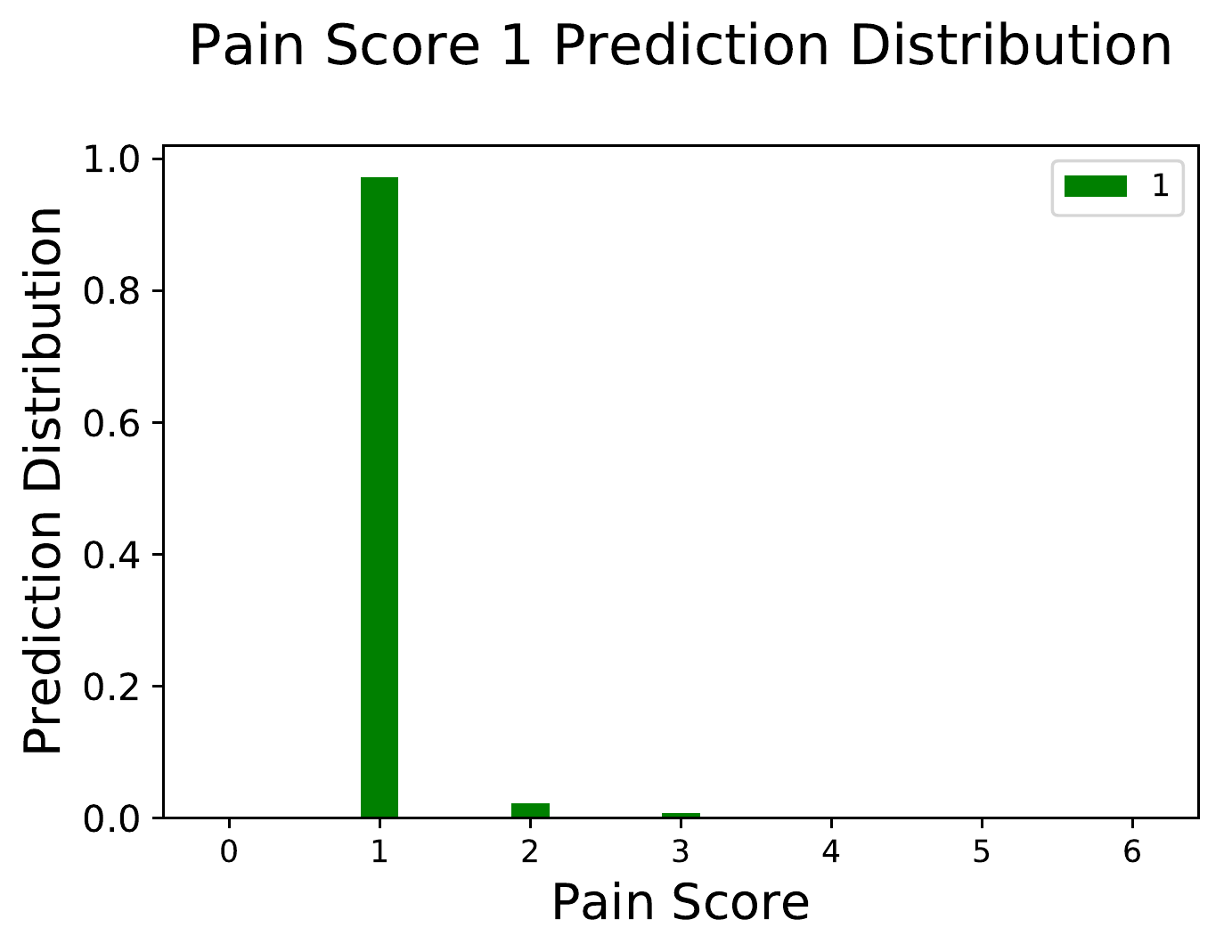}}
    \quad
    \subfloat{\includegraphics[width=0.23\textwidth]{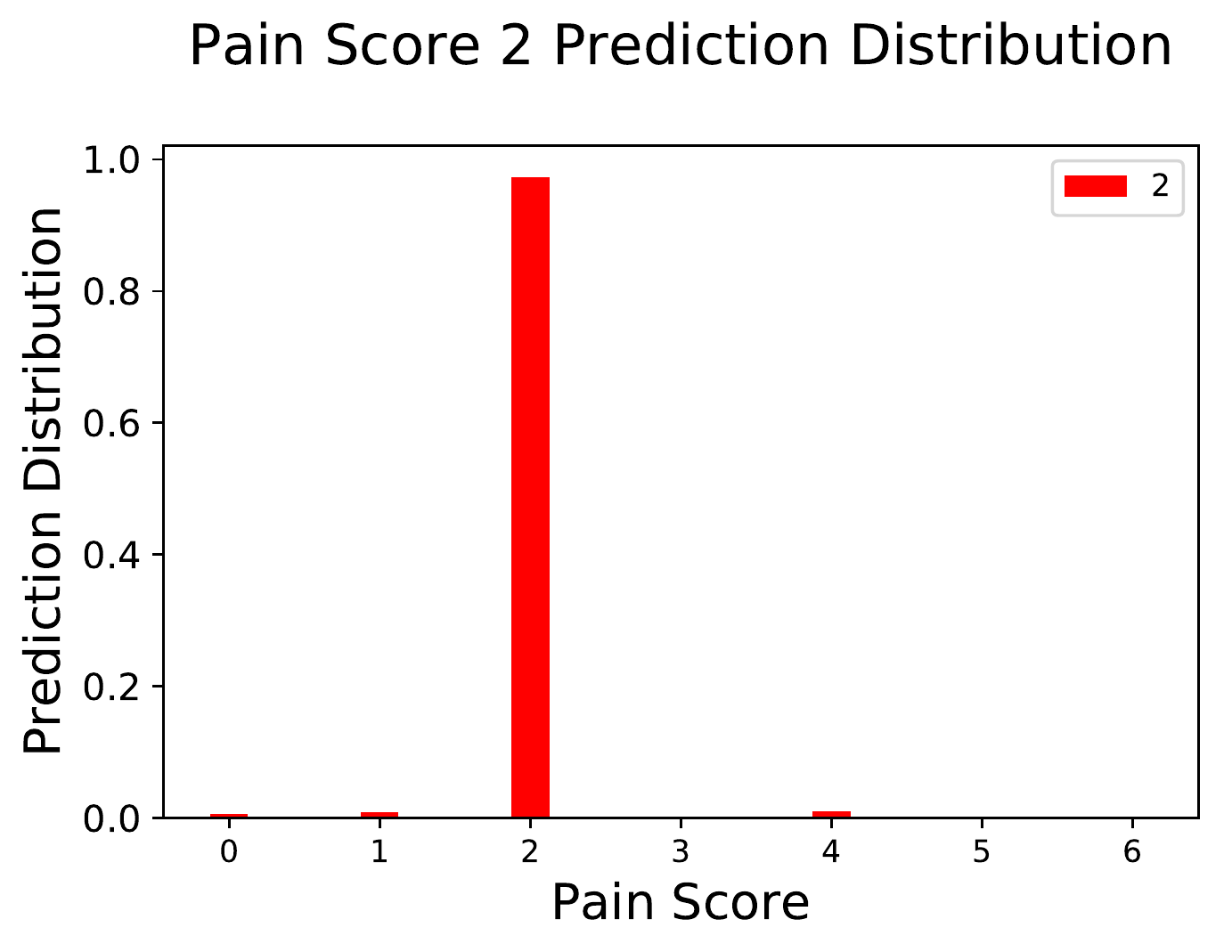}}
    \hfill
    \subfloat{\includegraphics[width=0.23\textwidth]{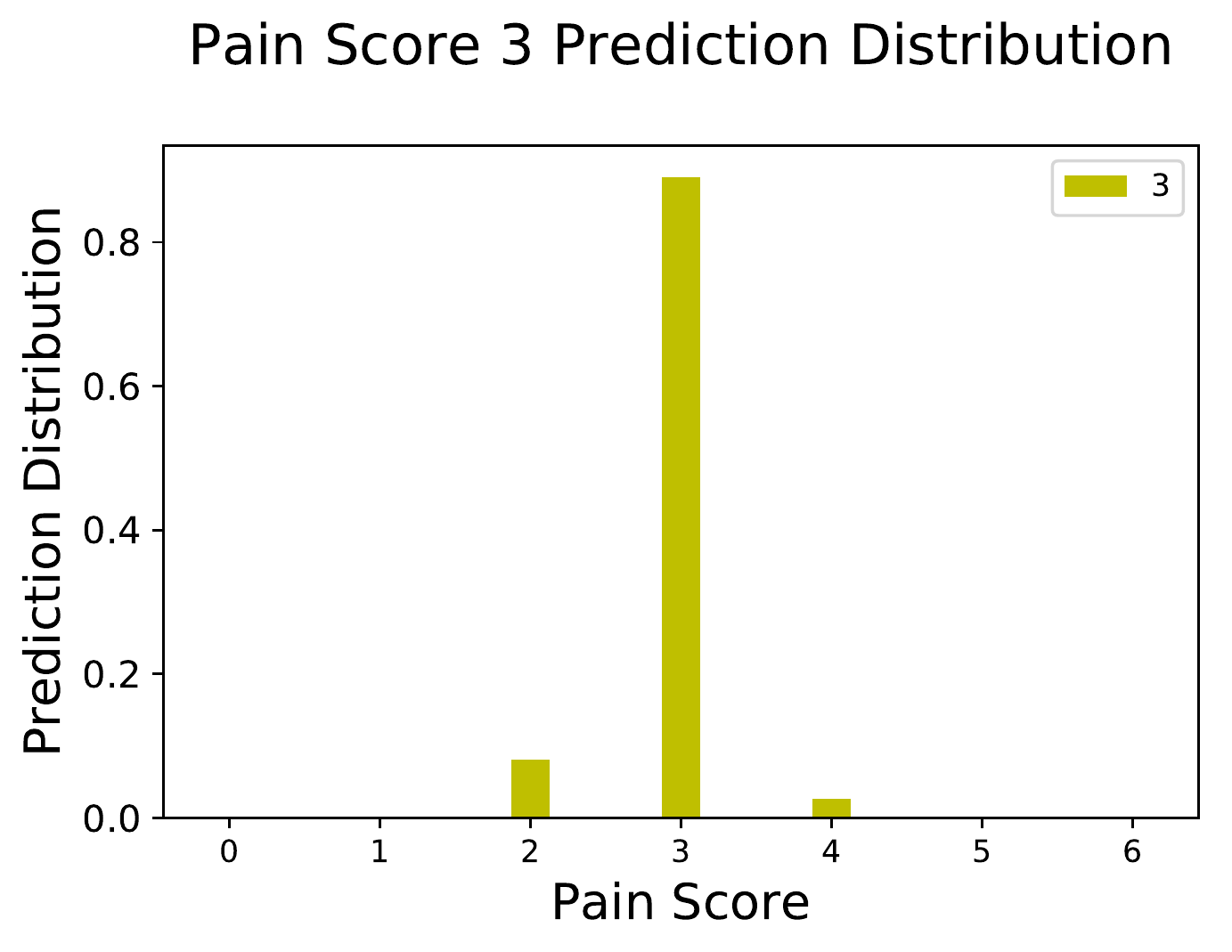}}
    \quad
    \subfloat{\includegraphics[width=0.23\textwidth]{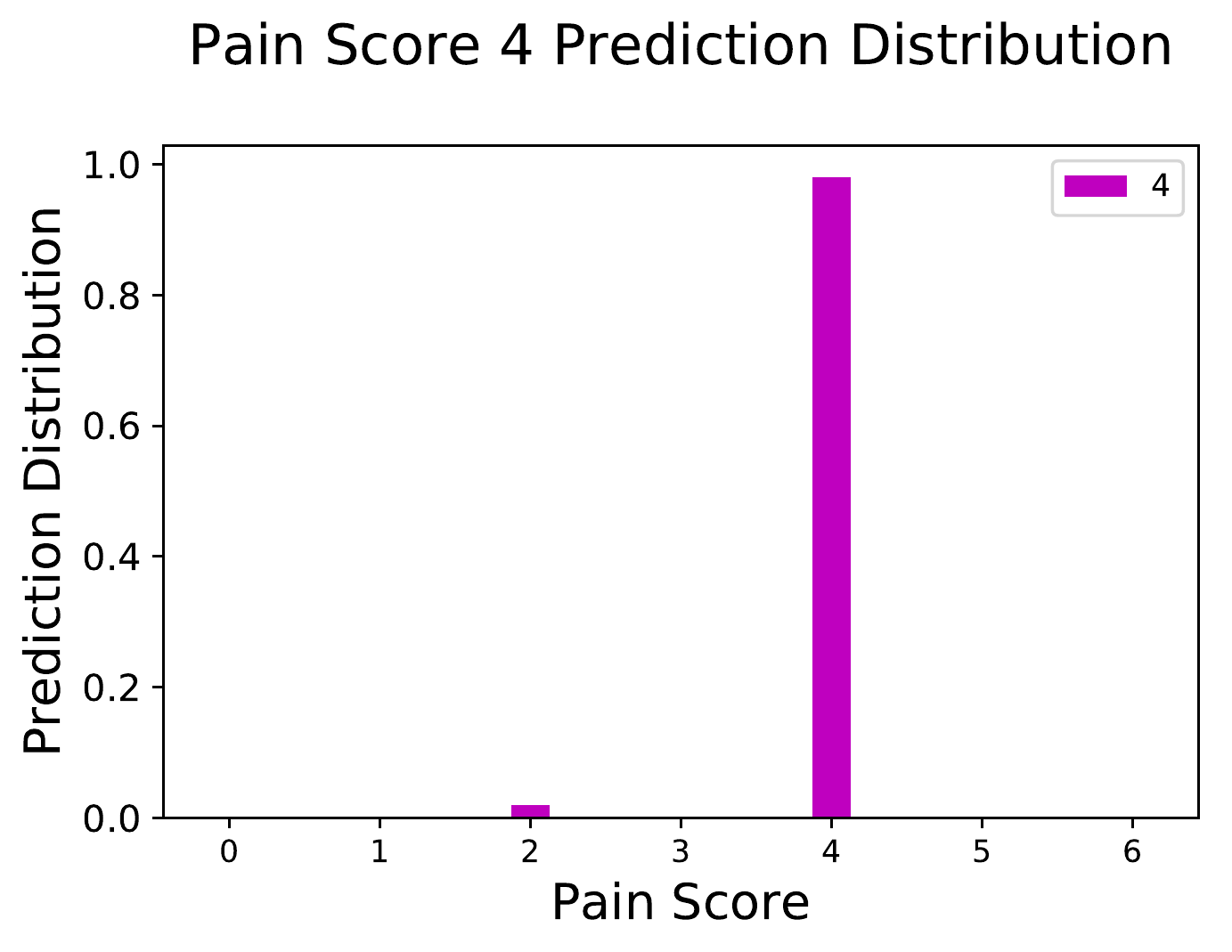}}
    \hfill
    \subfloat{\includegraphics[width=0.23\textwidth]{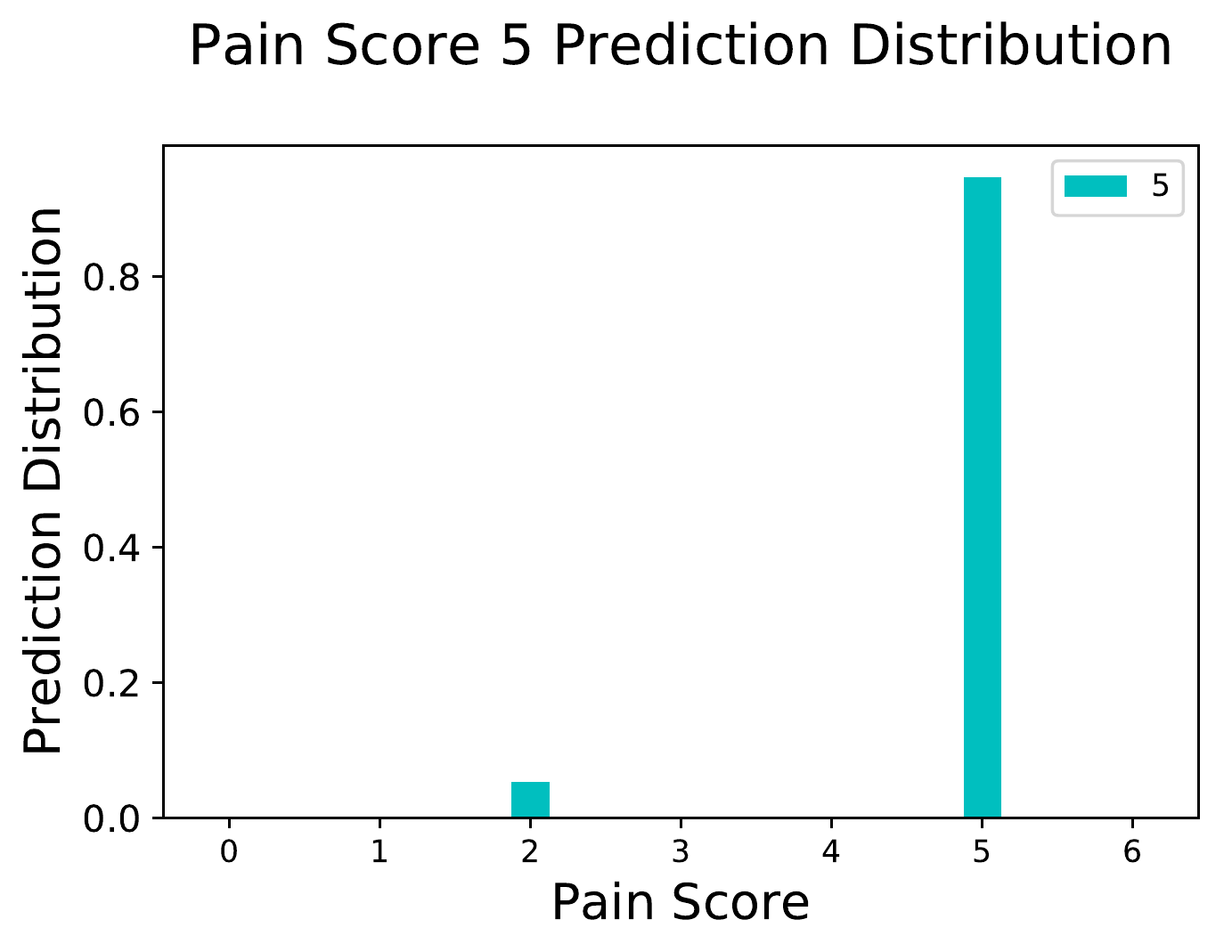}}
    \quad
    \subfloat{\includegraphics[width=0.23\textwidth]{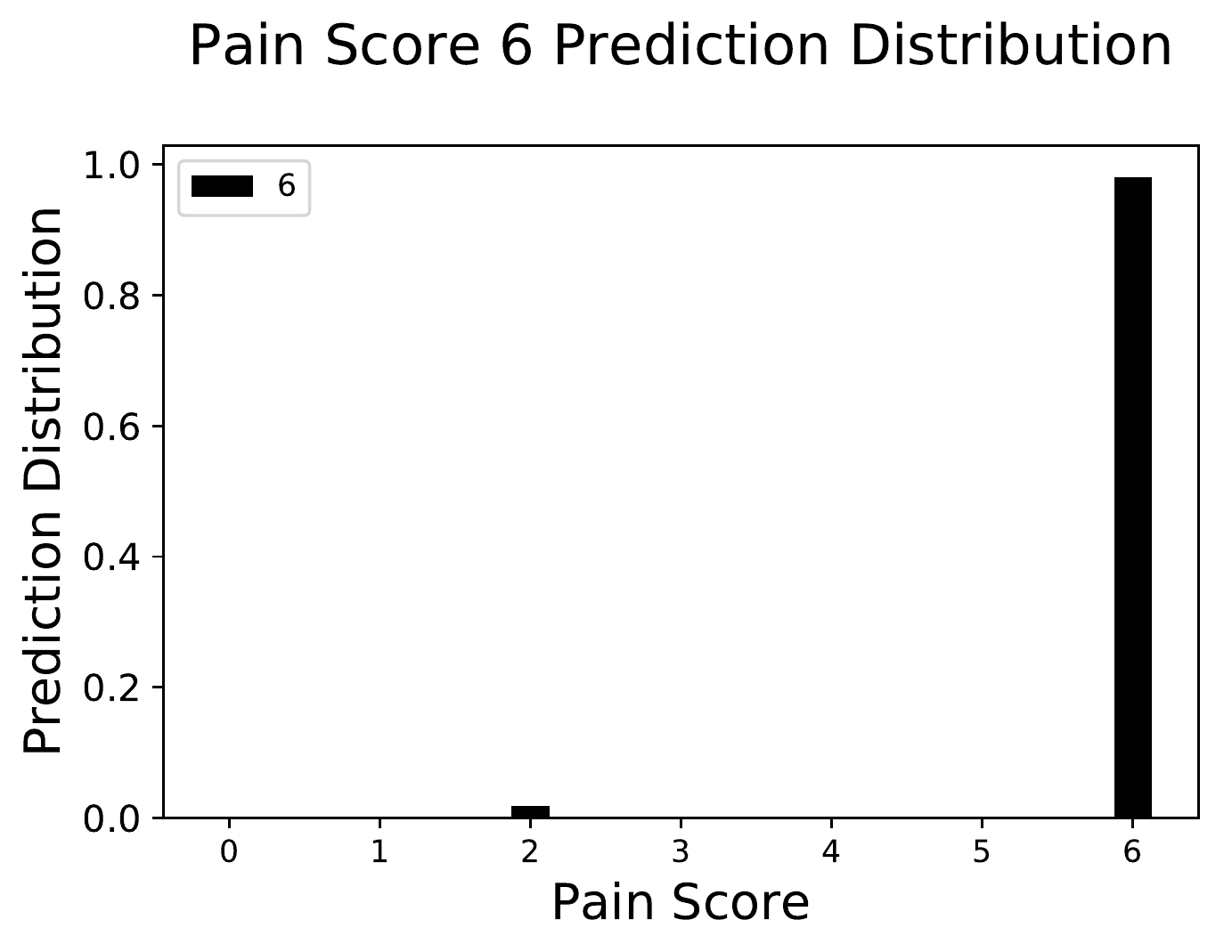}}
  \caption{Individual class prediction distribution for Dataset 2. The occasional incorrect predictions scatter close to the ground truth. }
  \label{fig:separate}
\end{figure}

\begin{figure}[ht]
  \includegraphics[width=\linewidth]{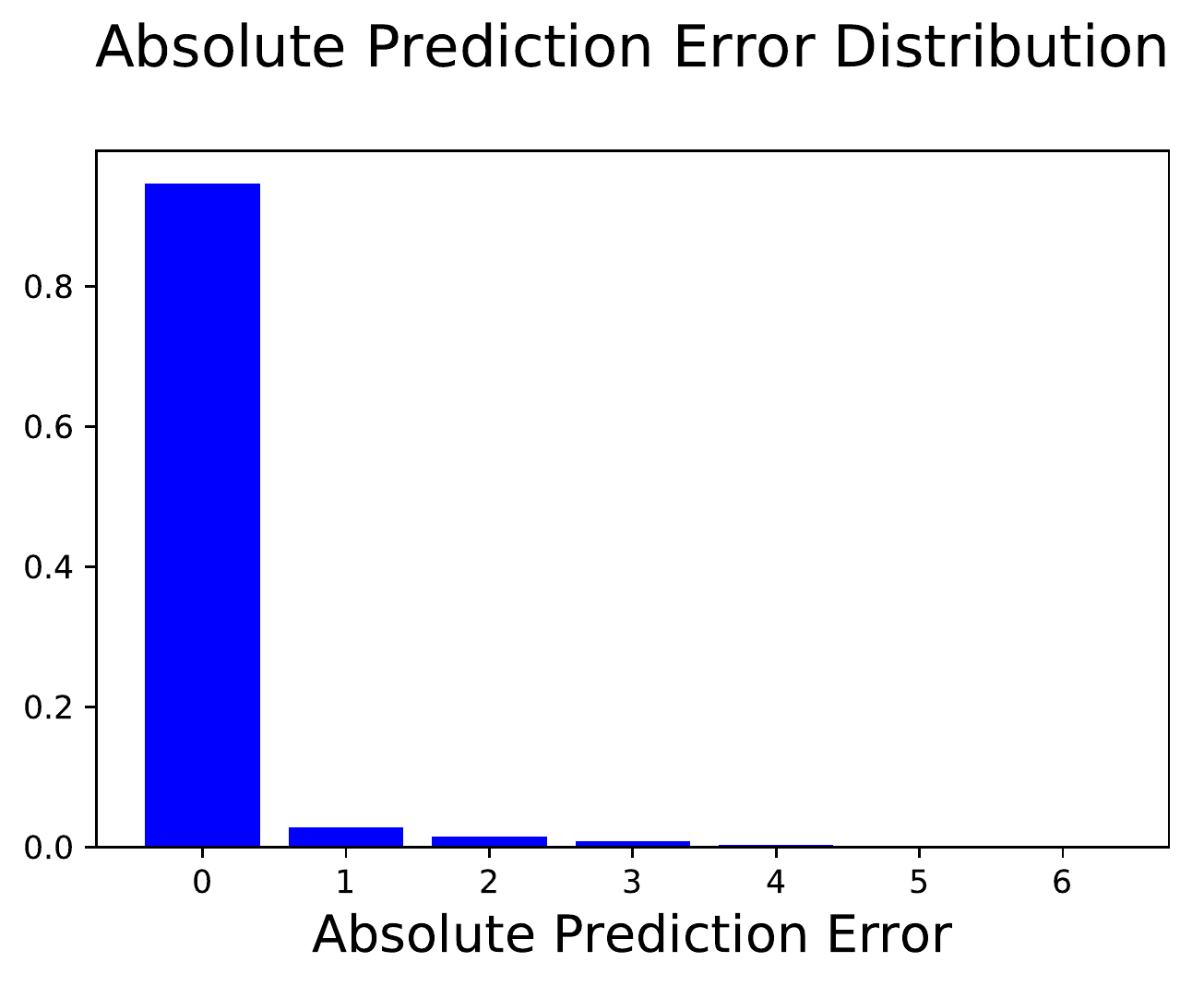}
  \caption{The distribution of absolute prediction error using ordinal regression for Dataset 2. The chance of making an error decreases with the increase of absolute prediction error.}
  \label{fig:APE}
\end{figure}

\begin{figure}[ht]
  \includegraphics[width=\linewidth]{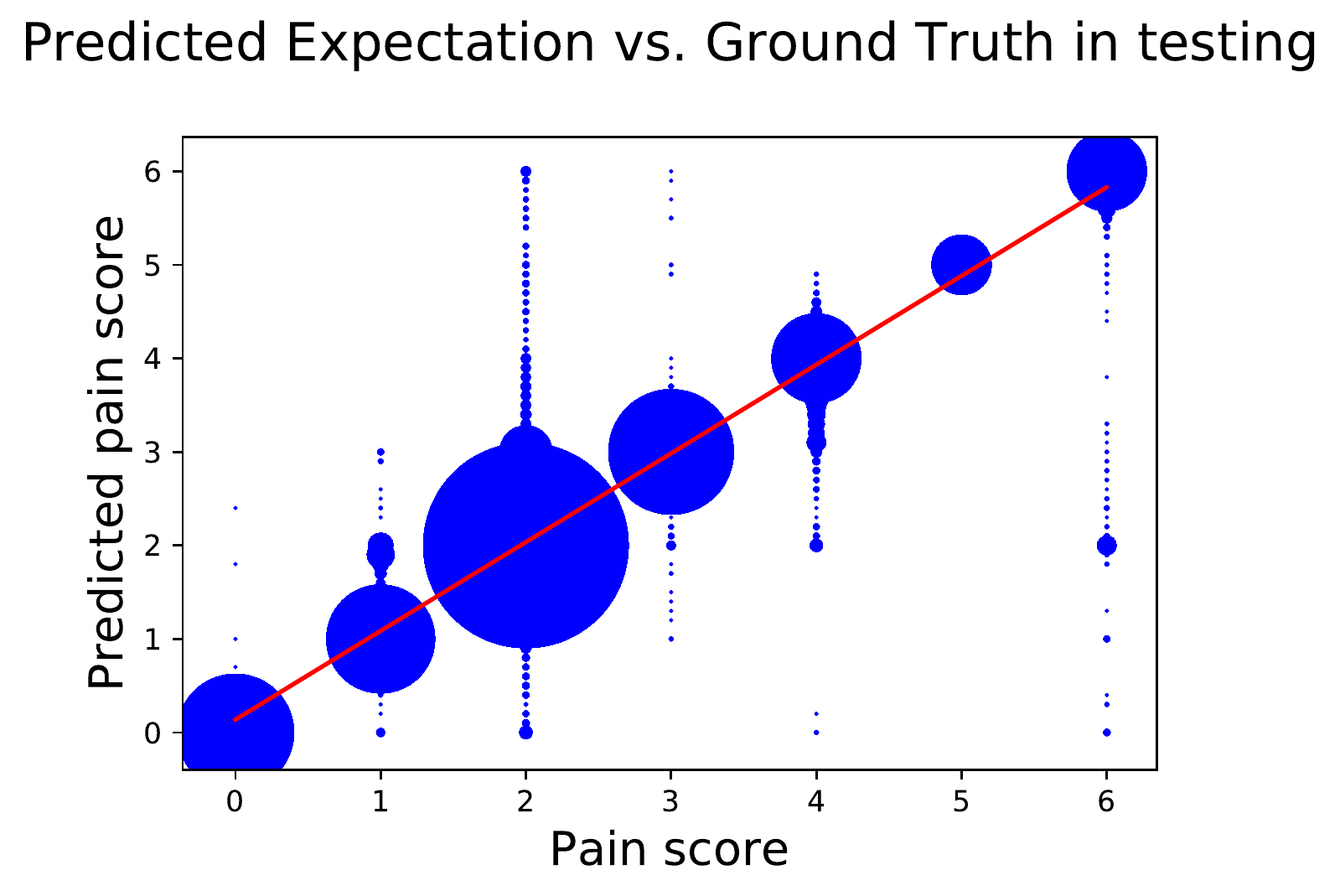}
  \caption{Scatter plot of expected predicted pain score vs. self-reported pain score for Dataset 2. The larger size of the point corresponds to the higher appearance frequency of the data point. We keep a decimal in calculating the expected predicted pain score because 0.1 is a reasonable precision to estimate pain in real life. The predicted pain score exerts high linear relationship with ground truth with a high $R^2$ of 0.9463.}
  \label{fig:regression}
\end{figure}

Confusion matrices are shown in Fig.~\ref{Confusion_Matrix_1} and Fig.~\ref{Confusion_Matrix_2} for chronic pain score prediction. 
Dominant numbers on the confusion matrix diagonal indicates that our model achieves high accuracy for each class. Fig.~\ref{fig:separate} shows the individual class  prediction  distribution  for  Dataset  2. 
The infrequent prediction mistakes scatter around the ground truth. For example, although 11\% of the predictions from pain score 3 are incorrect, they are still close to 3 (either 2 or 4). Fig~\ref{fig:APE} further demonstrates that the  probability  of  making  an  error  decreases as the absolute  prediction error increases. This is the benefit from ordinal regression, which penalizes more in the loss if the absolute error between ground truth and prediction is larger. Fig.~\ref{fig:regression} is a scatter plot of expected predicted pain score vs. self-reported pain score. The relationship between predicted pain score and ground truth is highly linear, with an R-squared ($R^2$) of 0.9463.

Since we use seq\_length to split long recordings into shorter slices, some of the short slices may not contain enough information for pain score prediction. However, these effects can be eliminated using Consensus Prediction. Although we  have  imblanced  data in Dataset 2,  we still use accuracy to compare different models and check the benefits obtained from Consensus Prediction, since chronic pain subjects  are   most  interested  in  prediction accuracy. 

\begin{figure}[ht]
  \includegraphics[width=\linewidth]{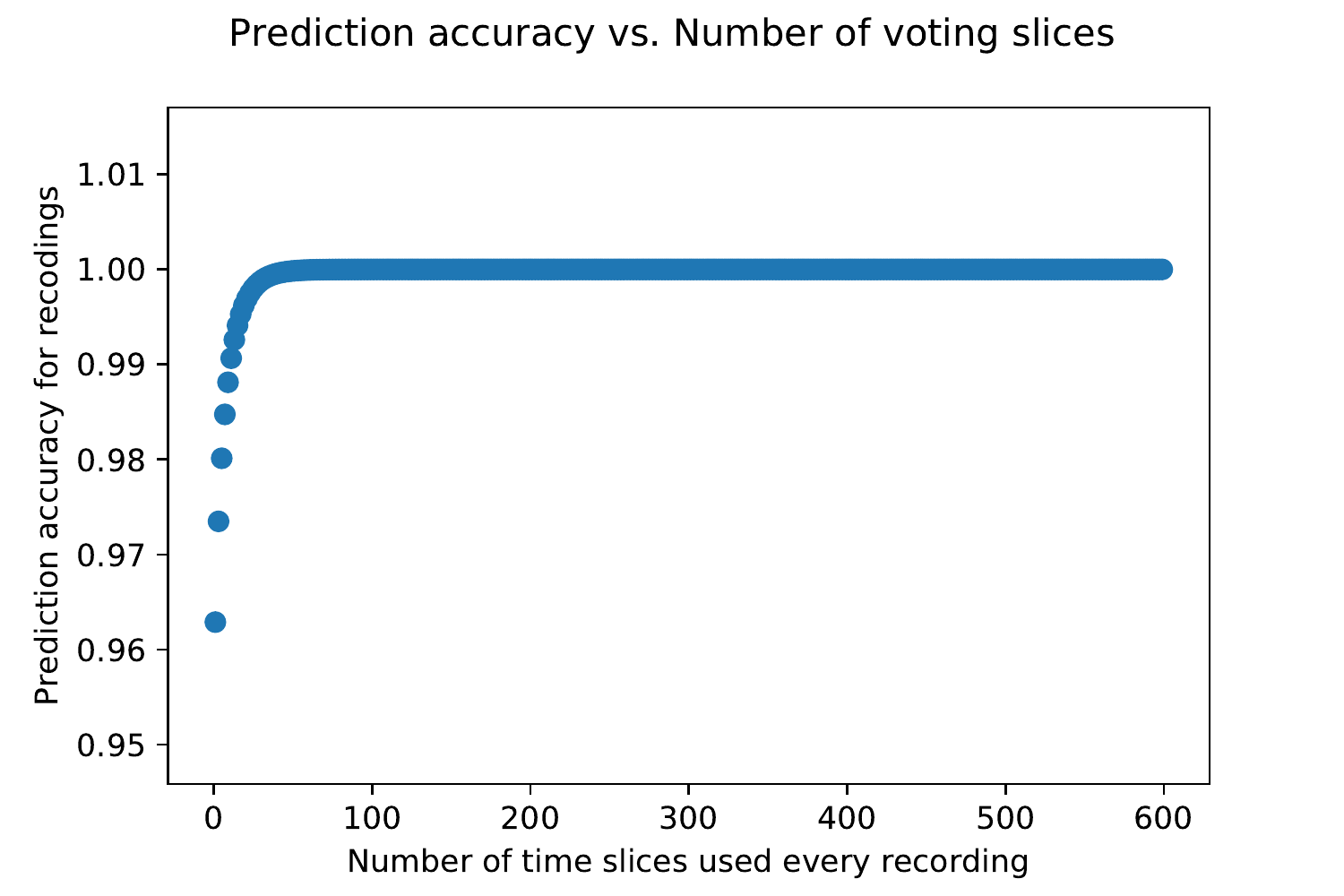}
  \caption{Consensus Prediction for chronic pain score prediction in Dataset 1.}
  \label{Pain_level}
\end{figure}

\begin{figure}[ht]
  \includegraphics[width=\linewidth]{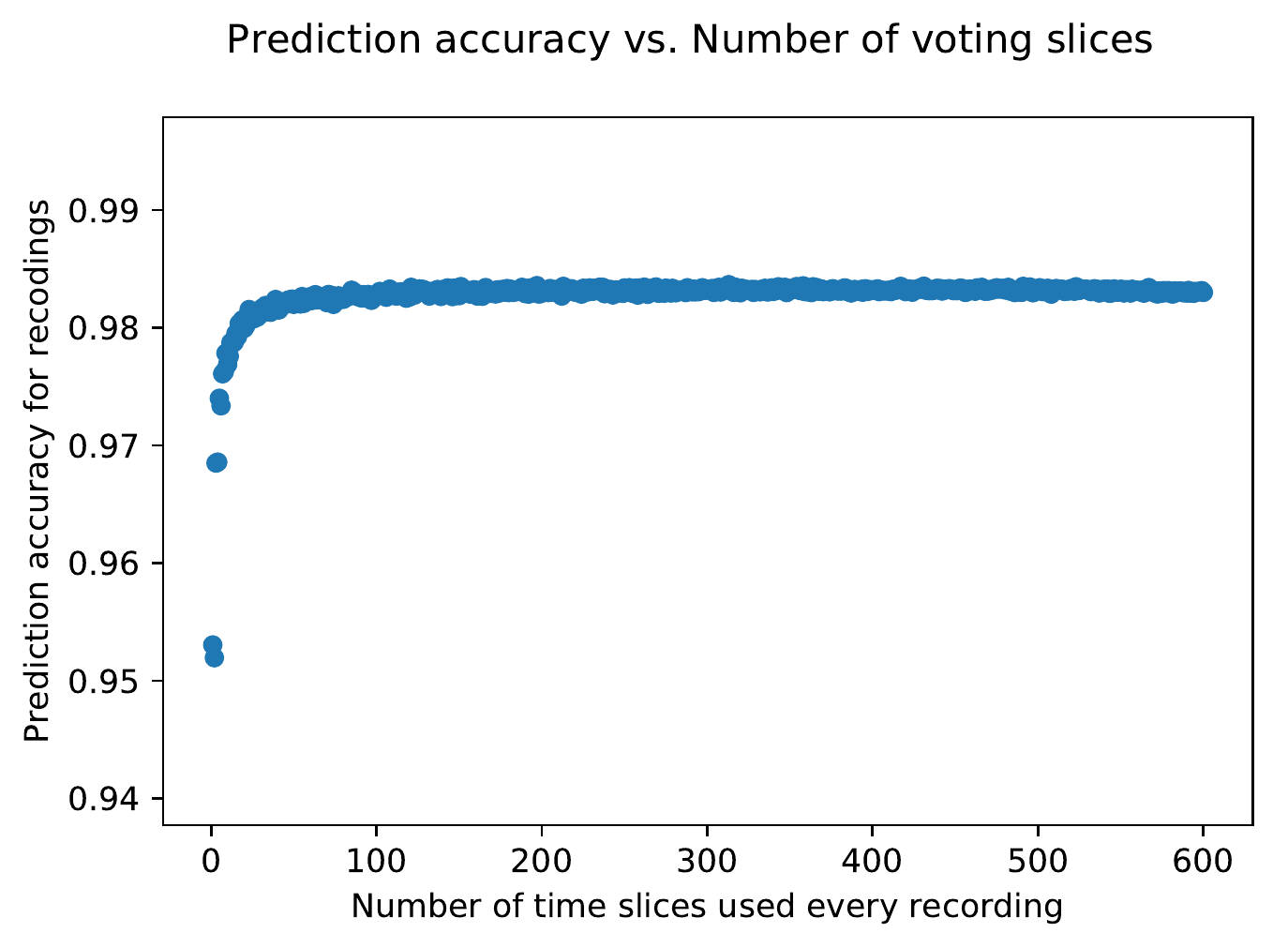}
  \caption{Consensus Prediction for chronic pain score prediction in Dataset 2.}
  \label{A_predl}
\end{figure}

Results of our framework compared against other machine learning models on chronic pain recordings are shown in Tables~\ref{perform_table_pain} and~\ref{perform_table_a} respectively. We compare our CNN based model with two baselines: Multilayer Perceptron (MLP) and feature-based Logistic Regression. For Dataset 1, our CNN-based deep learning approach improves the prediction accuracy by 5.25\% compared to feature-based Logistic Regression. Fig.~\ref{Pain_level} shows the Consensus Prediction accuracy. The accuracy improves by 3.84\% using Consensus Prediction. Although not all of the short slices can be expected to contain enough useful recording patterns, we can overcome that when we synthesize multiple individual classification results from these short slices. For Dataset 2, our model achieves accuracy of 95.23\% for short recording slices, which is a 2.8\% improvement over feature based Logistic Regression. The accuracy further improves to 98.30\% with Consensus Prediction as shown in Fig.~\ref{A_predl}. Our CNN based deep learning model also outperforms MLP on both of the two sets of recordings by 3.32\% and 2.91\% respectively, which shows CNN's advantage of local feature extraction using convolutional kernels over MLP. Also from Fig.~\ref{Pain_level} and Fig.~\ref{A_predl}, 100 time slices for each recording are sufficient in Consensus Prediction.

\begin{table}
\caption{Cross-validation performance comparison of our deep learning model with Multilayer Perceptrons and Logistic Regression on Dataset 1.}\label{perform_table_pain}
\begin{center}
\begin{tabular}{|l|l|}
\hline
Model &  Accuracy on Testing \\
\hline
\textbf{Convolutional Neural Network} &  \textbf{0.9630} \\
Multilayer Perceptron &  0.9321 \\
Logistic Regression & 0.9150\\

\hline
\end{tabular}
\end{center}
\end{table}

\begin{table}
\caption{Cross-validation  performance comparison of our deep learning model with Multilayer Perceptrons and Logistic Regression on Dataset 2. Although we have imblanced data for 7 classes, subjects are still most interested in prediction accuracy. We showed confusion matrix for our CNN model in the previous figure.}\label{perform_table_a}
\begin{center}
\begin{tabular}{|l|l|}
\hline
Model &  Accuracy on Testing \\
\hline
\textbf{Convolutional Neural Network} &  \textbf{0.9523} \\
Multilayer Perceptron &  0.9238 \\
Logistic Regression & 0.9063\\

\hline
\end{tabular}
\end{center}
\end{table}

\begin{figure}[ht]
\centering
\includegraphics[width=0.8\linewidth]{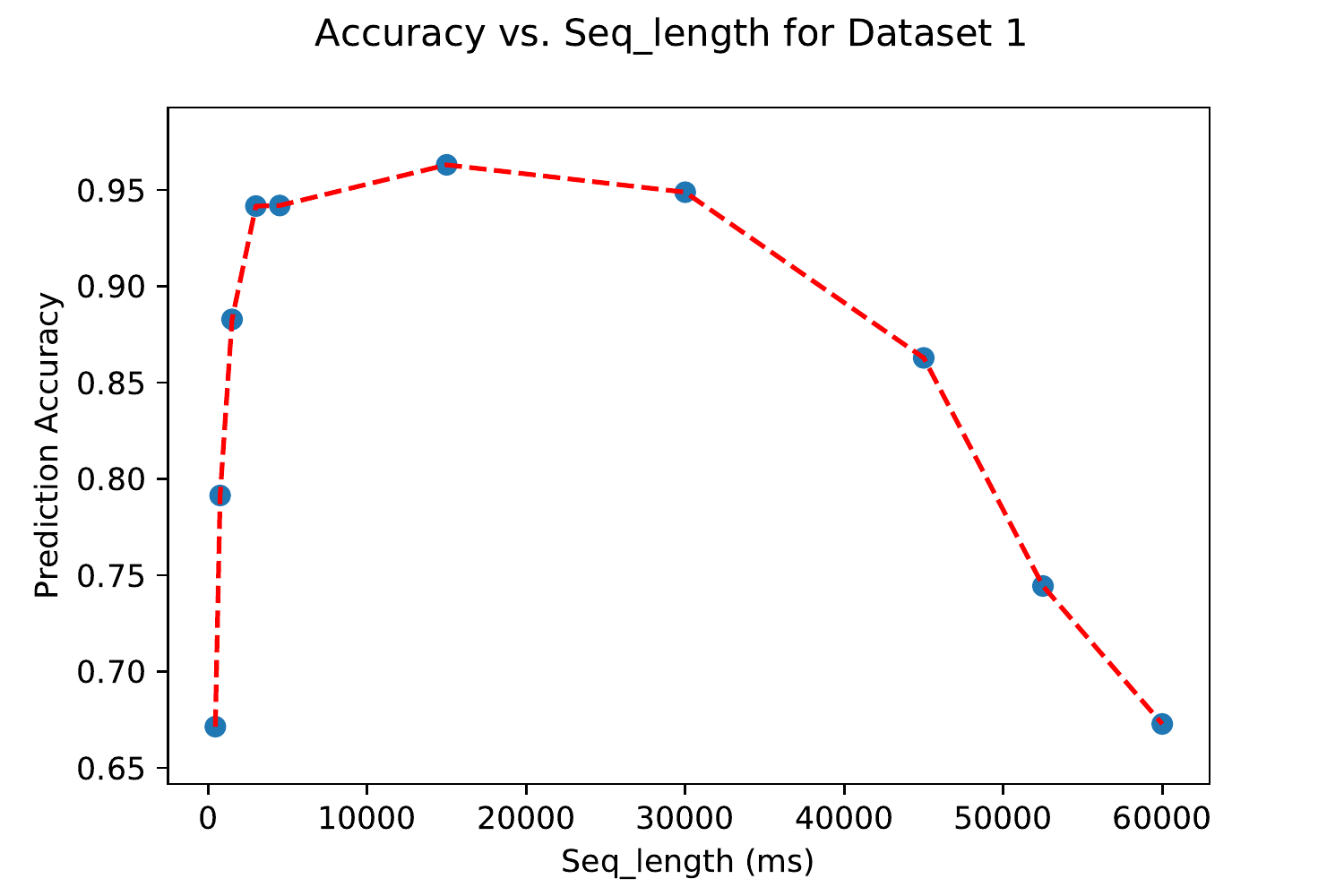}
\caption{Accuracy vs. Seq\_length trend for Dataset 1.} \label{seq_length}
\end{figure}

Fig.~\ref{seq_length} shows the trend of accuracy versus the choice of seq\_length for Dataset 1. For the effect of seq\_length on accuracy, there exists a trade-off between number of training samples and representation of a whole recording. The short slices contain less information but can provide more independent training samples. For deep learning models, larger numbers of training slices help more than a larger sample. However, we still cannot choose too small of a seq\_length, since a too short slice is not representative for a recording. Given the data we currently have, we use a seq\_length of 15 seconds.

\begin{figure}[ht]
\centering
\includegraphics[width=0.8\linewidth]{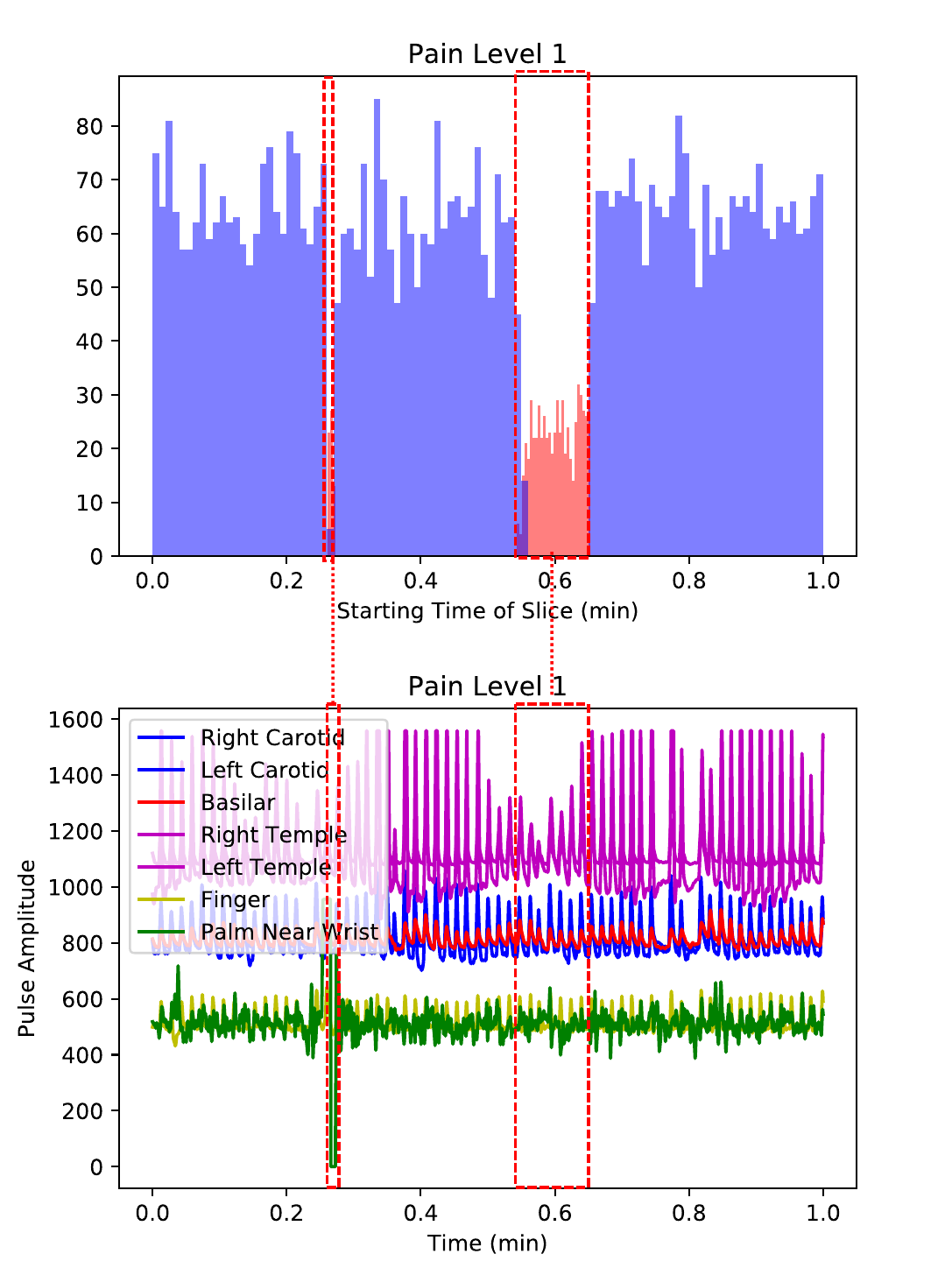}
\caption{Case study for short slices that the classifier predicted incorrectly. In the upper figure, red color bars represent incorrect predictions while blue bars represent correct predictions.} \label{case_study}
\end{figure}

We show some failure cases for our CNN classifiers in Fig.~\ref{case_study}. The upper figure shows the histogram of short sequence prediction for one recording, where blue indicates predicted successfully and red indicates predicted incorrectly. The x axis is the starting point of the sampled short sequences. The lower figure shows the corresponding voltage signals from our prototype Pain Meter recording. There is a period from 0.55 minute to 0.65 minute, during which the classifier consistently makes incorrect predictions. From the pulse signals, they also perform abnormally compared to other periods. This is also true for the short peak in the data coming from the palm near wrist at 0.28 minute. This illustrates that for those periods, by looking at shorter sequences, it is easy to make a wrong prediction for both human experts and a classifier. This is due to the fact that our sensors are sensitive to movements. Noise can be introduced with a contact position change between sensors and skin. However, by using Consensus Prediction, these errors have no effect on the final prediction.

\begin{figure}[ht]
  \includegraphics[width=\linewidth]{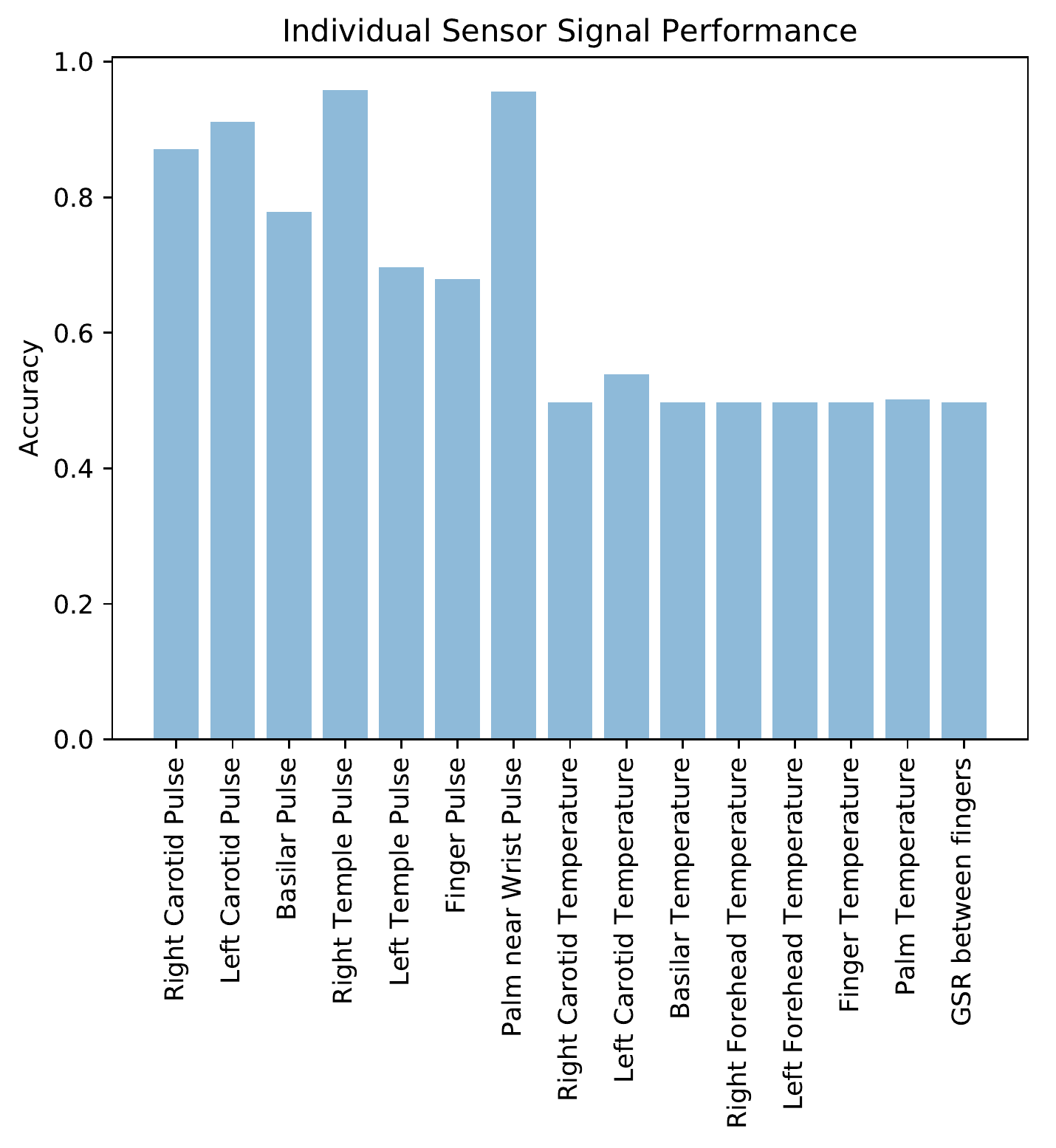}
  \caption{Individual sensor signal classification performance for Dataset 1.}
  \label{individuall}
\end{figure}

\section{Optimizing the Pain Meter Design}\label{optimize}
One of the benefits of our deep learning framework is that it can distinguish which sensors are most useful for pain score assessment. We compare the performance of each of the separate signals in Pain Meter 1 as shown in Fig.~\ref{individuall}. We report the average accuracy using 5-fold cross validation. This can help us to further improve our chronic Pain Meter. By testing individual signal performance, we find that the accuracy for temperature signals are all around 0.5, which is similar to a random guess in binary classification. Thus the temperature signals are not very informative. We also find that the temple pulse and the palm near wrist pulse contribute substantially to our pain score prediction.

\section{Discussion}\label{Discuss}
\begin{figure}[ht]
  \includegraphics[width=\linewidth]{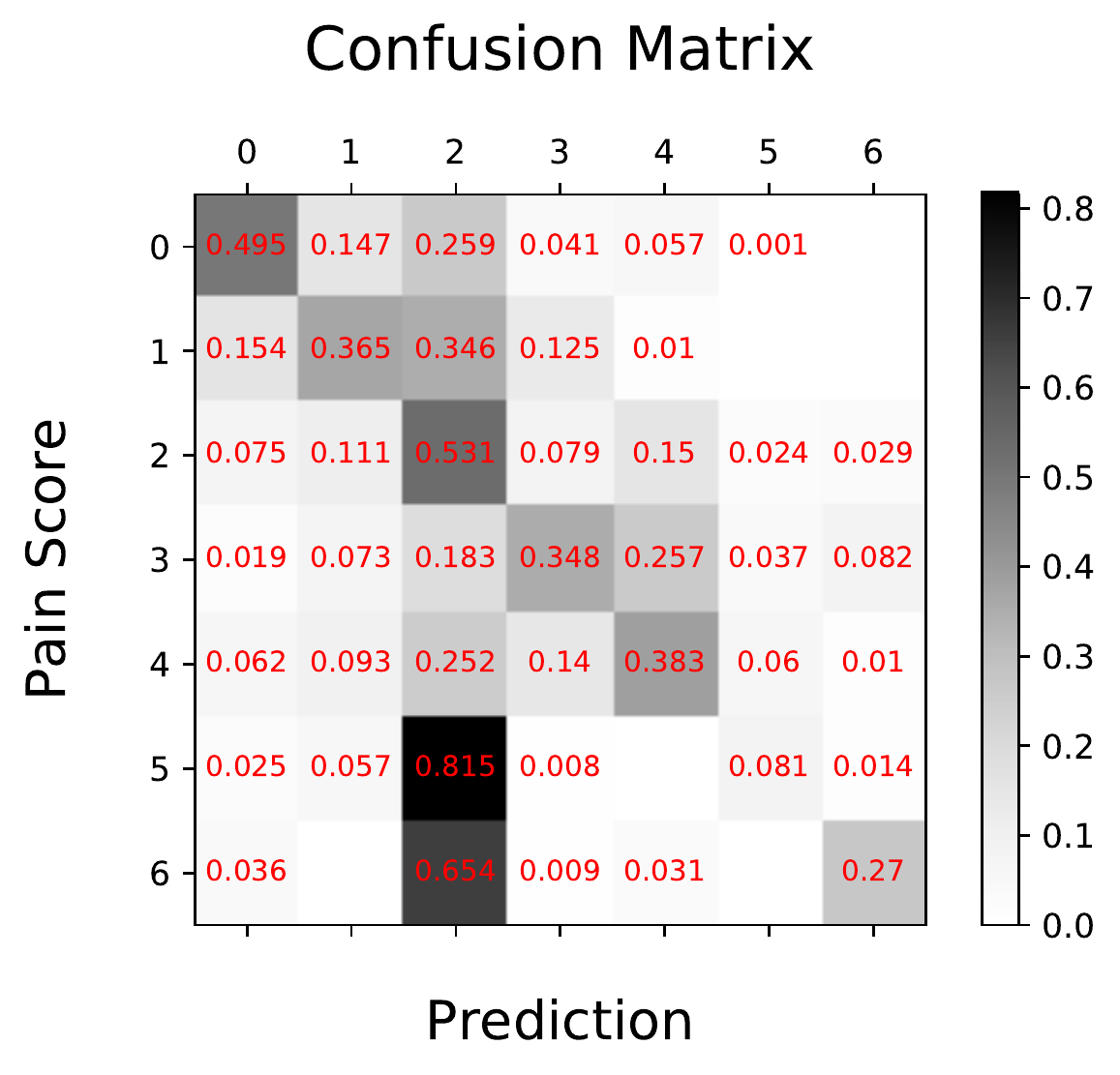}
  \caption{Confusion matrix for chronic pain score prediction in Dataset 2 using leave one recording out. The performance is poor due to subjects' differing sensitivities to pain, the use of different settings while recording and the limited unbalanced dataset.}
  \label{Confusion_Matrix_bad}
\end{figure}

We have addressed the issue of predicting a chronic pain score by proposing a deep learning  ordinal  regression framework. We split the long recordings into smaller slices, which not only eases the burden on GPU memory but also provides more training samples for the deep learning model. We define and use Consensus Prediction during testing. 
We present the Confusion Matrix of leave-one-recording-out in Fig.~\ref{Confusion_Matrix_bad}. 
Leave-one-recording-out does not perform well due to subjects' differing perceptions of pain intensities.
Also since we have much more pain score 2 samples compared with other pain score samples in Dataset 2, as is shown in Fig.~\ref{fig:statistics}, the model is prone to make predictions of pain score 2.

This paper is a proof of principle for chronic pain score assessment via deep learning. It can provide an objective pain assessment for each patient.  More data for additional chronic pain subjects is needed before it can be definitively known if deep learning will be a generally useful technique for chronic pain assessment.


\section{Acknowledgments}

 We thank the chronic pain subjects who donated their time for our data collection.

\bibliographystyle{IEEEtran}
\bibliography{references}

\begin{thebibliography}{10}
\providecommand{\url}[1]{#1}
\csname url@samestyle\endcsname
\providecommand{\newblock}{\relax}
\providecommand{\bibinfo}[2]{#2}
\providecommand{\BIBentrySTDinterwordspacing}{\spaceskip=0pt\relax}
\providecommand{\BIBentryALTinterwordstretchfactor}{4}
\providecommand{\BIBentryALTinterwordspacing}{\spaceskip=\fontdimen2\font plus
\BIBentryALTinterwordstretchfactor\fontdimen3\font minus
  \fontdimen4\font\relax}
\providecommand{\BIBforeignlanguage}[2]{{%
\expandafter\ifx\csname l@#1\endcsname\relax
\typeout{** WARNING: IEEEtran.bst: No hyphenation pattern has been}%
\typeout{** loaded for the language `#1'. Using the pattern for}%
\typeout{** the default language instead.}%
\else
\language=\csname l@#1\endcsname
\fi
#2}}
\providecommand{\BIBdecl}{\relax}
\BIBdecl

\bibitem{Imagenet}
A.~Krizhevsky, I.~Sutskever, and G.~E. Hinton, ``Imagenet classification with
  deep convolutional neural networks,'' in \emph{Advances in neural information
  processing systems}, 2012, pp. 1097--1105.

\bibitem{NLP}
R.~Collobert and J.~Weston, ``A unified architecture for natural language
  processing: Deep neural networks with multitask learning,'' in
  \emph{Proceedings of the 25th international conference on Machine
  learning}.\hskip 1em plus 0.5em minus 0.4em\relax ACM, 2008, pp. 160--167.

\bibitem{Speech}
A.~Graves, A.-r. Mohamed, and G.~Hinton, ``Speech recognition with deep
  recurrent neural networks,'' in \emph{2013 IEEE international conference on
  acoustics, speech and signal processing}.\hskip 1em plus 0.5em minus
  0.4em\relax IEEE, 2013, pp. 6645--6649.

\bibitem{Alphago}
D.~Silver, J.~Schrittwieser, K.~Simonyan, I.~Antonoglou, A.~Huang, A.~Guez,
  T.~Hubert, L.~Baker, M.~Lai, A.~Bolton \emph{et~al.}, ``Mastering the game of
  go without human knowledge,'' \emph{Nature}, vol. 550, no. 7676, p. 354,
  2017.

\bibitem{avati2018improving}
A.~Avati, K.~Jung, S.~Harman, L.~Downing, A.~Ng, and N.~H. Shah, ``Improving
  palliative care with deep learning,'' \emph{BMC medical informatics and
  decision making}, vol.~18, no.~4, p. 122, 2018.

\bibitem{dahlhamer2018prevalence}
J.~Dahlhamer, J.~Lucas, C.~Zelaya, R.~Nahin, S.~Mackey, L.~DeBar, R.~Kerns,
  M.~Von~Korff, L.~Porter, and C.~Helmick, ``Prevalence of chronic pain and
  high-impact chronic pain among adults—united states, 2016,''
  \emph{Morbidity and Mortality Weekly Report}, vol.~67, no.~36, p. 1001, 2018.

\bibitem{gaskin2012economic}
D.~J. Gaskin and P.~Richard, ``The economic costs of pain in the united
  states,'' \emph{The Journal of Pain}, vol.~13, no.~8, pp. 715--724, 2012.

\bibitem{van2019neuroimaging}
M.~M. van~der Miesen, M.~A. Lindquist, and T.~D. Wager, ``Neuroimaging-based
  biomarkers for pain: state of the field and current directions,'' \emph{Pain
  reports}, vol.~4, no.~4, 2019.

\bibitem{reddan2019brain}
M.~C. Reddan and T.~D. Wager, ``Brain systems at the intersection of chronic
  pain and self-regulation,'' \emph{Neuroscience letters}, vol. 702, pp.
  24--33, 2019.

\bibitem{reddan2018modeling}
------, ``Modeling pain using fmri: from regions to biomarkers,''
  \emph{Neuroscience bulletin}, vol.~34, no.~1, pp. 208--215, 2018.

\bibitem{cowen2015assessing}
R.~Cowen, M.~K. Stasiowska, H.~Laycock, and C.~Bantel, ``Assessing pain
  objectively: the use of physiological markers,'' \emph{Anaesthesia}, vol.~70,
  no.~7, pp. 828--847, 2015.

\bibitem{yang2018postoperative}
Y.~L. Yang, H.~S. Seok, G.-J. Noh, B.-M. Choi, and H.~Shin, ``Postoperative
  pain assessment indices based on photoplethysmography waveform analysis,''
  \emph{Frontiers in physiology}, vol.~9, p. 1199, 2018.

\bibitem{prichep2018exploration}
L.~S. Prichep, J.~Shah, H.~Merkin, and E.~M. Hiesiger, ``Exploration of the
  pathophysiology of chronic pain using quantitative eeg source localization,''
  \emph{Clinical EEG and neuroscience}, vol.~49, no.~2, pp. 103--113, 2018.

\bibitem{liu2012review}
H.~Liu, Y.~Wang, and L.~Wang, ``A review of non-contact, low-cost physiological
  information measurement based on photoplethysmographic imaging,'' in
  \emph{2012 Annual International Conference of the IEEE Engineering in
  Medicine and Biology Society}.\hskip 1em plus 0.5em minus 0.4em\relax IEEE,
  2012, pp. 2088--2091.

\bibitem{bargh2008unconscious}
J.~A. Bargh and E.~Morsella, ``The unconscious mind,'' \emph{Perspectives on
  psychological science}, vol.~3, no.~1, pp. 73--79, 2008.

\bibitem{kanawade2019photoplethysmography}
R.~Kanawade, S.~Tewary, H.~Sardana \emph{et~al.}, ``Photoplethysmography based
  arrhythmia detection and classification,'' in \emph{2019 6th International
  Conference on Signal Processing and Integrated Networks (SPIN)}.\hskip 1em
  plus 0.5em minus 0.4em\relax IEEE, 2019, pp. 944--948.

\bibitem{poplin2018prediction}
R.~Poplin, A.~V. Varadarajan, K.~Blumer, Y.~Liu, M.~V. McConnell, G.~S.
  Corrado, L.~Peng, and D.~R. Webster, ``Prediction of cardiovascular risk
  factors from retinal fundus photographs via deep learning,'' \emph{Nature
  Biomedical Engineering}, vol.~2, no.~3, p. 158, 2018.

\bibitem{darabi2018forecasting}
H.~R. Darabi, D.~Tsinis, K.~Zecchini, W.~F. Whitcomb, and A.~Liss,
  ``Forecasting mortality risk for patients admitted to intensive care units
  using machine learning,'' \emph{Procedia Computer Science}, vol. 140, pp.
  306--313, 2018.

\bibitem{jin2018treatment}
B.~Jin, H.~Yang, L.~Sun, C.~Liu, Y.~Qu, and J.~Tong, ``A treatment engine by
  predicting next-period prescriptions,'' in \emph{Proceedings of the 24th ACM
  SIGKDD International Conference on Knowledge Discovery \& Data Mining}.\hskip
  1em plus 0.5em minus 0.4em\relax ACM, 2018, pp. 1608--1616.

\bibitem{hsieh2018artificial}
M.-H. Hsieh, M.-J. Hsieh, C.-M. Chen, C.-C. Hsieh, C.-M. Chao, and C.-C. Lai,
  ``An artificial neural network model for predicting successful extubation in
  intensive care units,'' \emph{Journal of clinical medicine}, vol.~7, no.~9,
  p. 240, 2018.

\bibitem{miotto2017deep}
R.~Miotto, F.~Wang, S.~Wang, X.~Jiang, and J.~T. Dudley, ``Deep learning for
  healthcare: review, opportunities and challenges,'' \emph{Briefings in
  bioinformatics}, vol.~19, no.~6, pp. 1236--1246, 2017.

\bibitem{rodriguez2017deep}
P.~Rodriguez, G.~Cucurull, J.~Gonz{\`a}lez, J.~M. Gonfaus, K.~Nasrollahi, T.~B.
  Moeslund, and F.~X. Roca, ``Deep pain: Exploiting long short-term memory
  networks for facial expression classification,'' \emph{IEEE transactions on
  cybernetics}, 2017.

\bibitem{liu2017deepfacelift}
D.~Liu, F.~Peng, A.~Shea, R.~Picard \emph{et~al.}, ``Deepfacelift:
  interpretable personalized models for automatic estimation of self-reported
  pain,'' \emph{arXiv preprint arXiv:1708.04670}, 2017.

\bibitem{chu2017physiological}
Y.~Chu, X.~Zhao, J.~Han, and Y.~Su, ``Physiological signal-based method for
  measurement of pain intensity,'' \emph{Frontiers in neuroscience}, vol.~11,
  p. 279, 2017.

\bibitem{lotsch2018machine}
J.~L{\"o}tsch and A.~Ultsch, ``Machine learning in pain research,''
  \emph{Pain}, vol. 159, no.~4, p. 623, 2018.

\bibitem{pitcher2019prevalence}
M.~H. Pitcher, M.~Von~Korff, M.~C. Bushnell, and L.~Porter, ``Prevalence and
  profile of high-impact chronic pain in the united states,'' \emph{The Journal
  of Pain}, vol.~20, no.~2, pp. 146--160, 2019.

\bibitem{johnson2016machine}
A.~E. Johnson, M.~M. Ghassemi, S.~Nemati, K.~E. Niehaus, D.~A. Clifton, and
  G.~D. Clifford, ``Machine learning and decision support in critical care,''
  \emph{Proceedings of the IEEE. Institute of Electrical and Electronics
  Engineers}, vol. 104, no.~2, p. 444, 2016.

\bibitem{hira2015review}
Z.~M. Hira and D.~F. Gillies, ``A review of feature selection and feature
  extraction methods applied on microarray data,'' \emph{Advances in
  bioinformatics}, vol. 2015, 2015.

\bibitem{ben2013monitoring}
N.~Ben-Israel, M.~Kliger, G.~Zuckerman, Y.~Katz, and R.~Edry, ``Monitoring the
  nociception level: a multi-parameter approach,'' \emph{Journal of clinical
  monitoring and computing}, vol.~27, no.~6, pp. 659--668, 2013.

\bibitem{rubins2019imaging}
U.~Rubins, Z.~Marcinkevics, I.~Logina, A.~Grabovskis, and E.~Kviesis-Kipge,
  ``Imaging photoplethysmography for assessment of chronic pain patients,'' in
  \emph{Optical Diagnostics and Sensing XIX: Toward Point-of-Care Diagnostics},
  vol. 10885.\hskip 1em plus 0.5em minus 0.4em\relax International Society for
  Optics and Photonics, 2019, p. 1088508.

\bibitem{johansson1999monitoring}
A.~Johansson, P.~{\AA}. {\"O}berg, and G.~Sedin, ``Monitoring of heart and
  respiratory rates in newborn infants using a new photoplethysmographic
  technique,'' \emph{Journal of clinical monitoring and computing}, vol.~15,
  no. 7-8, pp. 461--467, 1999.

\bibitem{bonissi2013preliminary}
A.~Bonissi, R.~D. Labati, L.~Perico, R.~Sassi, F.~Scotti, and L.~Sparagino, ``A
  preliminary study on continuous authentication methods for
  photoplethysmographic biometrics,'' in \emph{2013 IEEE Workshop on Biometric
  Measurements and Systems for Security and Medical Applications}.\hskip 1em
  plus 0.5em minus 0.4em\relax IEEE, 2013, pp. 28--33.

\bibitem{PPG_link}
\url{pulseSensor.com} Accessed May 26, 2020.

\bibitem{zuzarte2019quantifying}
I.~Zuzarte, P.~Indic, D.~Sternad, and D.~Paydarfar, ``Quantifying movement in
  preterm infants using photoplethysmography,'' \emph{Annals of biomedical
  engineering}, vol.~47, no.~2, pp. 646--658, 2019.

\bibitem{BatchNorm}
S.~Ioffe and C.~Szegedy, ``Batch normalization: Accelerating deep network
  training by reducing internal covariate shift,'' \emph{arXiv preprint
  arXiv:1502.03167}, 2015.

\bibitem{Dropout}
N.~Srivastava, G.~Hinton, A.~Krizhevsky, I.~Sutskever, and R.~Salakhutdinov,
  ``Dropout: a simple way to prevent neural networks from overfitting,''
  \emph{The journal of machine learning research}, vol.~15, no.~1, pp.
  1929--1958, 2014.

\bibitem{Earlystop}
L.~Prechelt, ``Early stopping-but when?'' in \emph{Neural Networks: Tricks of
  the trade}.\hskip 1em plus 0.5em minus 0.4em\relax Springer, 1998, pp.
  55--69.

\bibitem{Adam}
D.~P. Kingma and J.~Ba, ``Adam: A method for stochastic optimization,''
  \emph{arXiv preprint arXiv:1412.6980}, 2014.

\end{thebibliography}

\end{document}